\documentclass[preprint]{aastex}
\usepackage{amsmath,mathrsfs}
\usepackage[dvips]{color}
\usepackage{natbib}
\newcommand{\ms}{M$_{\odot}$}

\newcommand{\pd}[2] { { {\partial #1}\over{\partial #2}}}

\newcommand{\mpy}{M$_{\odot}$ yr$^{-1}$}
\newcommand{\macc}{$\dot{\rm M}_{acc}$} 
\begin{document}
\title{ The Impact of Dust Evolution and Photoevaporation on Disk Dispersal }
\author{U.Gorti\altaffilmark{1,2} , D.~Hollenbach\altaffilmark{1} \& C.~P.~Dullemond\altaffilmark{3}}
\altaffiltext{1} {SETI Institute, Mountain View, CA}
\altaffiltext{2} {NASA Ames Research Center, Moffett Field, CA} 
\altaffiltext{3} {University of Heidelberg, Germany}    

\begin{abstract}
Protoplanetary disks are dispersed by viscous evolution and photoevaporation in a few million years; in the interim small, sub-micron sized dust grains must grow and form planets. The time-varying abundance of small grains in an evolving disk directly affects gas heating by far-ultraviolet photons, while dust evolution affects photoevaporation by changing the disk opacity and resulting penetration of FUV photons in the disk. Photoevaporative flows, in turn, selectively carry small dust grains leaving the larger particles---which decouple from the gas---behind in the disk. We study these effects by investigating the evolution of a disk subject to viscosity, photoevaporation by EUV, FUV and X-rays, dust evolution, and radial drift using a 1-D multi-fluid approach (gas + different dust grain sizes) to solve for the evolving surface density distributions. The 1-D evolution is augmented by 1+1D models constructed at each epoch to obtain the instantaneous disk structure and determine  photoevaporation rates. The implementation of a dust coagulation/fragmentation model results in a marginal decrease in disk lifetimes  when compared to models with no dust evolution; the disk lifetime is thus found to be relatively insensitive to the evolving dust opacity. We find that photoevaporation can cause significant reductions in the gas/dust mass ratio in the planet-forming regions of the disk as it evolves, and may result in a corresponding increase in heavy element abundances relative to hydrogen. We discuss  implications for theories of planetesimal formation and giant planet formation, including the formation of gas-poor giants.  After gas disk dispersal, $\sim 3\times 10^{-4}$ \ms\ of mass in solids typically remain, comparable to the solids inventory of our solar system.
\end{abstract}

\section{Introduction}
The lifetime of a circumstellar disk sets the timescale for planet formation which must compete with disk dispersal for raw material.  Studies of gas and dust both point to a characteristic lifetime for disks $\lesssim 10$ Myrs (e.g., Zuckerman et al. 1995, Haisch et al. 2001, Pascucci et al. 2006, Hernandez et al. 2007). The main dispersal mechanisms of disks are viscous evolution and photoevaporation. Viscous accretion spreads disk mass, causes accretion onto the central star, and is primarily responsible for getting rid of the innermost ($r\lesssim 1$AU) matter well within the stellar gravitational potential well.  Photoevaporation  results from thermal heating of gas on the disk surface that drives mass loss flows and ultimately removes outer ($r\gtrsim$1AU) disk material. Timescales derived from disk dispersal theories  depend on the agent responsible for photoevaporation ---  stellar EUV, FUV or X-ray photons --- and on disk viscosity.  Disk dispersal times calculated using theoretical models are reasonably concurrent with observationally estimated disk lifetimes $\sim 1-5$ Myrs (e.g., Alexander et al. 2006, Owen et al. 2010, 2012, Gorti, Dullemond \& Hollenbach 2009, hereafter GDH09). While most of the gas is dispersed due to viscosity and photoevaporation, a significant fraction of the dust may perhaps be incorporated into planets (e.g., Chiang \& Youdin 2010, Chiang \& Laughlin 2013, Hansen \& Murray 2013). 

The formation of Solar System planets, the diversity of observed exoplanets, and the giant planet-stellar metallicity correlation all support the core accretion paradigm of planet formation (Pollack et al. 1996) which begins with the collisional growth of small sub-micron sized dust particles in the disk (Johansen et al. 2014).  Gravity aids the subsequent formation of planetesimals, Earth-sized rocky masses, and ultimately the cores of the giant planets. Current theories estimate timescales of $\sim 3 $Myr for the formation of cores that are massive enough to accrete gas and form giant planets (e.g., Lissauer et al. 2009). 
 
Observations of dust in disks confirm the first step in the core accretion hypothesis; there are  clear indications of dust evolution as disks age (see review by Testi et al. 2014).  Sub-micron sized interstellar dust grains are inferred to grow to at least millimeter sizes in disks from the flattening of the opacity law ($\kappa_{\nu} \propto \nu^{\beta}$) at sub-millimeter wavelengths where dust is optically thin.  In  the ISM,  $\beta \sim 1.7$, whereas in protoplanetary disks, $\beta \sim 0.9-1.0$ (Beckwith et al. 1990), indicating coagulation of grains and a subsequent reduction in dust opacity.  Further, disks have  been imaged at millimeter and centimeter wavelengths (e.g., Andrews et al. 2012, Isella et al. 2013) confirming the presence of large particles.  Although dust evolution all the way to planetary sizes is uncertain, there appears to be convincing evidence that grains grow to at least $\sim$ cm sizes at early epochs in disk evolution.
 
Several mechanisms for dust grain growth have been proposed with the earliest models indicating a highly efficient process (e.g, Weidenschilling \& Cuzzi 1993, Dullemond \& Dominik 2005) with almost all dust being removed on short timescales $\sim 10^5$ years. More recent models highlight barriers to growth (e.g., the fragmentation barrier, Brauer et al. 2008; bouncing barrier, Zsom et al. 2010) limiting the maximum sizes attained to $\lesssim 0.1-1$cm. Introducing a more realistic Maxwellian distribution of collision velocities is found to allow the unobstructed growth of a few objects, while the bulk of the collisions still retard growth due to fragmentation and keep particle sizes within $\sim$ 1cm or less (Windmark et al. 2012, Garaud et al. 2013 and Laibe 2014). As we discuss in this paper, the time evolution of the disk further impedes the growth of grains as the declining gas surface density increases dust collision speeds and promotes shattering. 
 
Grain growth reduces the number of smaller, sub-micron grains and affects disk heating. The vertical structure of the disk is determined by the  $\sim$ sub-$\mu$m sized dust particles due to their capacity to absorb stellar photons. In the simple model of vertical hydrostatic equilibrium which  describes the bulk disk structure reasonably well, a surface layer of dust intercepts the stellar photons and cools by radiation in the infrared (e.g., Chiang \& Goldreich 1997). These infrared photons then penetrate and heat dust in the optically thick midplane where gas and dust are collisionally well coupled.  The properties of the dust in the surface layers are thus critical in determining the midplane temperature and the collisional coupling of gas and dust at the surface, and in setting the structure of the disk.  The small particles dominate even though their individual cross sections are smaller;  collectively they are so much more numerous that they mainly determine the opacity at visible wavelengths. Gas at the disk surface is additionally heated by the grain photoelectric mechanism acting on small grains due to FUV photons.  Grain growth to centimeter sizes implies a drop in the dust opacity (per H nucleus) in the disk by orders of magnitude at optical and ultraviolet wavelengths. And importantly, the column density of gas and dust in these surface layers determines the penetration of UV and X-ray photons that heat the gas and cause photoevaporation. 

Photoevaporation results due to heating of gas by high energy photons (FUV, EUV and X-rays) that drives thermal winds.  For FUV-driven photoevaporation, mass loss rates are determined not only by the strength of the radiation field but also by the abundance of small grains and PAH (polycyclic aromatic hydrocarbon) molecules that are responsible for grain photoelectric heating (e.g., Weingartner \& Draine 2001).  The column density of FUV heated gas  strongly depends on the penetration depth and hence on the small dust population that attenuates FUV photons. Dust evolution may thus hold consequences for  how the disk loses mass. Conversely, photoevaporation affects dust evolution. Typically, grain size distributions are such that most of the mass of the grains is in the large particles; but the photoevaporative flow will only carry away small dust grains that are collisionally coupled to gas. Hence, photoevaporation can lower the gas to dust mass ratio in the disk and also influence subsequent dust evolution.

Diminished gas/dust ratios create conditions favorable for planetesimal formation---the second step towards building planets in the core accretion scenario. Gravitational instabilities (Goldreich \& Ward 1973), sedimentation and sweep-up (Youdin \& Shu 2002), streaming instabilities (Youdin \& Johansen 2007, Bai \& Stone 2010) and turbulent concentrations (Cuzzi et al. 2008, 2010) are all mechanisms that have been proposed to circumvent the difficulties in growing cm-sized objects to km-sizes. All these mechanisms are facilitated by reductions in the gas/dust ratio.  

Due to the implications of the radial distribution of solids in gaseous disks  for planet formation, the global evolution of dust has been studied with interest. Stepinski \& Valageas (1996) first investigated the relative motion of gas and solids as solid particles decouple from the gas. In their analysis, where they assumed the presence of grains of different sizes, they found that only the smallest ($\lesssim$ cm) particles are coupled to the midplane gas leading to substantial changes in the radial distribution: an enhancement in the solid surface density at $r\lesssim10$ AU, and a depletion of solids relative to gas beyond. Takeuchi, Clarke \& Lin (2005) combined a viscous gas disk evolution model with EUV photoevaporation and a model for the evolution of the dust. They studied the evolution of mm-size grains over the lifetime of the disk, and found that  gas drag causes them to migrate inwards on $\sim$ Myr timescales. They concluded that drag causes all dust to spiral into the star and that disks must finally evolve to a gas-rich state. Alexander \& Armitage (2007) conducted a similar analysis, with EUV photoevaporation (no dust evolution and a single particle size) that included the creation of a gap at late stages. Their focus was on the late stages of evolution when the formation of an inner hole and irradiation of the rim leads to rapid disk dispersal. The creation of a gap was found to result in a pile up of dust near the rim. Hughes \& Armitage (2012) re-examined this problem, again with EUV photoevaporation, but with a grain growth formulation. They found that the outer disk becomes depleted in solids, as did Stepinski \& Valageas (1996), but that there is no radial pile-up of material in the inner disk.  Instead the gas/dust ratio remains roughly the same for most of the disk lifetime because of radial advection. EUV photoevaporation acting at late stages led to reductions in the gas/dust mass ratio, but the time to disperse the disk after these epochs was so short and the dust mass remaining so small that there was no significance for planet formation. They concluded that global disk evolution models do not provide the required reductions in gas/dust ratios that could lead to rapid planetesimal formation (e.g., Goldreich \& Ward 1973, Youdin \& Shu 2002, Cuzzi et al 2008, 2010, Bai \& Stone 2010).

This paper revisits this topic but differs from the above earlier work, primarily due to the additional consideration of FUV and X-ray photoevaporation and a more sophisticated dust evolution model.  As discussed in the preceding paragraphs, dust evolution must have consequences for photoevaporation in this context, especially due to the changing penetration depth of FUV photons and grain photoelectric heating.  Here we present the first study of the impact of dust evolution on disk dispersal via viscous evolution and photoevaporation by stellar EUV, FUV and X-ray photons.  We use previously developed 1-D models of disk evolution and  in this paper add dust evolution following a recent analysis by Birnstiel, Ormel \& Dullemond (2011, hereafter BOD11) of the dust size distribution of grains subject to coagulation, fragmentation, settling and turbulence. Further, our models include a 1+1D calculation of the disk in order to determine the vertical structure of the disk for computing the FUV/X-ray photoevaporation rates.   While previous models assume all dust is left behind by photoevaporation, we explicitly calculate the fraction of dust that is carried away by the flow and that which is retained.  We examine the effect of dust evolution on photoevaporation rates and dispersal times, and investigate the evolution of the gas/dust mass ratio in the disk. 

The outline of the paper is as follows: \S 2 provides an overview of disk evolution and summarizes the main results of our earlier work on photoevaporation (GDH09) which did not consider dust evolution, \S 3 describes the additions to our disk model and our treatment of dust evolution, and \S 4 discusses the results. In \S 5, we consider the implications of this work on disk evolution and planet formation and end with a summary in \S6.

\section{Disk Evolution with Viscosity and Photoevaporation}
We first present a brief overview of disk evolution under the combined  influences of viscosity and photoevaporation and summarize the results of GDH09  (see also the review by Alexander et al.~2014). 

Our disk evolution models begin after the central star has accumulated most of its mass, $\sim0.5-1$ Myrs into its life through the Class 0/I  phases (e.g., Evans et al. 2009, Bell et al. 2013).  Prior to this, disk evolution is typically dominated by gravitational instabilities, vigorous accretion and strong jets and outflows.  We ignore this phase of evolution and start calculations  ($t=0$)  when the disk has become gravitationally stable,  with $M_{disk}=0.1$M$_*$, and when viscous accretion dominates transport of disk mass.  With time, the disk spreads to larger radii and also accretes mass  onto the star, lowering the surface density at each radius and thereby lowering the accretion rate onto the star.   Since the jet/outflow scales with accretion, it becomes weaker (e.g., Hartmann et al.~1998,  White \& Hillenbrand 2004) and therefore more transparent, allowing stellar high energy photons to penetrate  the outflow column and irradiate the disk. Penetration is first by hard X-ray\footnote[1]{Hard X-rays--$E\gtrsim$1keV; Soft X-rays--$\sim 0.1-0.3$keV; EUV--13.6eV-100eV; FUV-- 6-13.6eV} and FUV photons when accretion rates fall to \macc$<10^{-6}$ M$_{\odot}$ yr$^{-1}$ and hydrogen nuclei column densities through the base of the outflow are ${\rm N}_{\rm H}\sim 10^{22-24}$ cm$^{-2}$, and later by  EUV and soft X-ray  photons  (\macc$<10^{-8}$ M$_{\odot}$ yr$^{-1}$,  ${\rm N}_{\rm H}\sim 10^{19-20}$cm$^{-2}$) (GDH09). This surface irradiation  marks the onset of significant photoevaporation where gas at the surface is heated to temperatures high enough to overcome gravity and drive mass loss flows (e.g., Alexander et al. 2006, Ercolano et al. 2009, Gorti \& Hollenbach 2009, GDH09, Owen et al. 2010, 2012). 

The consequent evolution of the disk is determined by the decrease in surface density with time due to both viscous diffusion and photoevaporation. If photoevaporation dominates viscosity at any radial location, then diffusive processes are unable to redistribute mass rapidly enough and the disk clears at this radius.  This typically occurs in two places: (i) in the inner disk, at $\sim 1-10$AU, due to heating of gas to $\sim 1000-10^4$ K by EUV, FUV or X-ray photons (depending on their relative strengths) with the possible creation of a gap,  and (ii) in the outer disk at $\gtrsim 100-300$ AU where even gas temperatures of $\sim100$K can cause mass loss due to weakened gravity, truncating the disk here and halting its slow viscous expansion. Depending on the relative contributions of accretion and stellar chromospheric activity that lead to high energy photons, disks may thereafter evolve to either form gaps in the inner regions and disperse in an  ``inside-out" manner, or in some instances may disperse without forming gaps and shrink radially with time (see \S 4.1).

In GDH09, we concluded that for typical stellar high energy radiation fields with equal luminosities at X-ray, EUV and FUV wavelengths, FUV photons were primarily responsible for photoevaporation. At these levels, we found that EUV photoevaporation was not significant (cf. Alexander et al. 2014, Pascucci et al. 2014), and that X-ray photons were found to dominate disk mass loss only if the flux incident on the disk surface had a predominantly soft spectrum (also see Owen et al. 2010). In all cases, FUV photons were found to drive mass loss in the outer regions of the disks where most of its mass resides. A key feature of FUV photoevaporation is that the viscous expansion of the outer edge of the disk with time  is curbed (under the assumption that the viscous parameter $\alpha$ does not change with disk radius).  For typical disk parameters, we found that approximately half the mass is accreted onto the star while the remaining half the disk mass is lost due to photoevaporation.

Photoevaporation rates, and hence disk lifetimes, were determined to depend critically on the stellar radiation field which is a strong function of stellar mass.  With the assumption that the initial disk mass M$_{disk}\propto$ M$_*$, disks around low mass stars ($\lesssim 3$M$_{\odot}$) were found to live relatively longer ($\sim 3-6\times10^6$ years) in spite of their lower disk mass and strong FUV and X-rays due to accretion and chromospheric activity. More massive stars ($\gtrsim 3 $M$_{\odot}$) were found to disperse their disks fast (in $\lesssim 10^6$ years) due to their high intrinsic photospheric FUV (and EUV for O stars) fluxes. 

In GDH09, we considered dust to be of a single size with a gas/dust mass ratio of  100 and perfectly coupled to the gas. Moreover, the dust size was assumed to be constant in space and time. Since FUV photoevaporation was found to be significant in determining disk lifetimes and because the abundance of small dust grains affect both FUV attenuation and FUV heating of gas, here we consider the role of dust evolution as the disk disperses. In the rest of the paper, we explore the effects of dust evolution on photoevaporation rates, disk evolution,  and disk lifetimes. 

\section{Disk Models}
We use a radial, 1-D model for the time-dependent evolution of the disk surface density that combines the effects of viscosity and photoevaporation by EUV, FUV and X-rays as described in earlier work (GDH09) with some modifications. In these models, an initial prescribed surface density profile for a disk is evolved viscously with time, as determined by the diffusion equation (e.g., Lynden-Bell \& Pringle 1974) using a simple $\alpha$-parametrization for the viscosity. The standard surface density evolution equation is augmented by a sink term that describes the photoevaporative mass loss. Photoevaporation results in a decrease in surface density with time ($\dot{\Sigma}_{pe}(r,t)$) and therefore,
\begin{equation}
{{\partial \Sigma}\over{\partial t}} = {\frac{3}{r} }{{\partial}\over{\partial r}} \left(
\sqrt r {{\partial}\over{\partial r}} \left( \nu \Sigma \sqrt r \right) \right) - \dot{\Sigma}_{pe} (r,t)
\label{sdot}
\end{equation}
where $r$ is the radial co-ordinate, $\Sigma$ is the surface density and $\dot{\Sigma}_{pe}$ is the instantaneous photoevaporation rate at that radius (from both sides of the disk) due to EUV, FUV and X-ray heating. The viscosity  coefficient $\nu = \alpha c_s^2/\Omega_K$, where $c_s$ is the  isothermal sound speed computed from the temperature of the disk at the midplane and $\Omega_K$  is the Keplerian angular frequency. 

At every instant of time, the photoevaporation rate, $\dot{\Sigma}_{pe}$ due to FUV, EUV and X-rays is calculated.  The instantaneous accretion rate onto the star is used to calculate a time-dependent accretion-generated FUV component which is added to the stellar chromospheric UV and X-ray field. The accretion shock is assumed to be a black body at 9000K (Gullbring et al. 2000), and the FUV spectrum ($912-2000$\AA) is accordingly computed. This procedure yields a FUV flux that is $\sim$30\% higher than the observationally derived relation of Yang et al. (2012) for high accretion luminosities $\sim 0.1$L$_{\odot}$ and $\sim$70\% lower for lower accretion luminosities $\sim 0.001$L$_{\odot}$. However, both these relations do not include the Lyman $\alpha$ contribution to the FUV flux which is expected to be substantial (e.g., Bergin et al. 2003), and therefore our FUV flux estimate may be conservative by a factor of a few. We have modified our X-ray spectrum to a two-temperature plasma (with equal 1MK and 10MK component thermal bremsstrahlung fluxes, e.g., Skinner \& Guedel 2013), based on recent high resolution {\em Chandra} and {XMM-Newton} observations. This makes X-rays slightly more significant than in GDH09, but otherwise does not  affect the results here substantially. While the spectrum is uncertain and soft X-rays may be accretion driven, we assume the spectrum is constant in time.   The chromospheric FUV luminosity and the EUV luminosity are assumed to be of the same magnitude as the X-ray luminosity in all our models and these are also  constant in time.  We further consider a protostellar wind (e.g., X-wind) accompanying accretion and the associated attenuation of the stellar flux through the wind column when calculating the irradiation of the disk  by high energy photons (see \S 2 and GDH09). 

 The surface density profile ($\Sigma(r,t)$) is used to  construct  a  1+1D disk structure model iterated from a calculated vertical gas temperature distribution with the assumption of vertical hydrostatic equilibrium. The dust model is a  simple two-layer  approximation similar to that described in Chiang \& Goldreich (1997). Optical and high-energy photons from the star irradiate the disk surface to heat the dust and gas. Thermal balance yields the gas disk vertical structure, the density $n$ and gas temperature $T_{gas}$ as a function of spatial location $(r,z)$ at every instant of time. The photoevaporation rate  $\dot{\Sigma}_{pe}(r,t)$ is determined as described below from this instantaneous disk structure model  and used in Equation~\ref{sdot}, which is solved numerically using an implicit Crank-Nicholson scheme (see GDH09). 

We have slightly revised our determination of $\dot{\Sigma}_{pe}(r,t)$ from the 1+1D disk structure at each timestep. As in GDH09, we use the Adams et al (2004) approximation to evaluate the potential $\dot{\Sigma}_{pe}(r,t)$ from each height $z$ for a given $r$ and consider the flow to be launched from the height where $\dot{\Sigma}_{pe}(r,t)$  is maximum. The thermal speed $c_s$ attained by the gas determines whether or not the flow can escape gravity and become unbound, the criterion $c_s^2=GM_*/r_g$ thus defines a characteristic gravitational radius $r_g$ (e.g., Hollenbach et al. 1994). The angular momentum of the Keplerian gas disk provides additional support against gravity, and flows  from a rotating disk are unbound when the disk radius exceeds a critical radius $r_{crit}\sim r_g/8$ (Liffman et al. 2003, Adams et al. 2004). Note that since the gas temperature tends to rise with $z$, $r_{crit}$ and $r_g$ decrease with increasing $z$ at fixed $r$. We assume that flows are launched subsonically and that there is a smooth transition of the flow velocity across the sonic or critical point, i.e., there are no shocks that propagate into the disk.  Following the recommendations of Waters \& Proga (2012), we do not approximate the flow streamline from $r_{crit}$ to $r_g$ as in GDH09. Instead, we assume that the flow is sonic if $r>r_{crit}$. We also add a missing factor of $\exp(-1/2)$ to our earlier expressions.  Thus the photoevaporation rate is given by 
 \begin{equation}
\dot{\Sigma}_{pe} (r) =  \max_z \left[ \rho_{base}(r,z)\  \mathscr{M}(r,r_g)\ c_s(r,z) \right]
\label{perate}
\end{equation}
where $\mathscr{M}$ is the flow Mach number, with
\begin{align}
\mathscr{M}(r,r_g)&= \left(\frac{r_s}{r}\right)^2 
 \exp\left[-\frac{1}{2} -\left(\frac{r_g}{2r}\right) \left(1-\frac{r}{r_s}\right)^2\right]  
 & \quad \text{for} \quad r<r_{crit}\quad\text{and} \notag \\ 
&= 1  & \qquad \text{otherwise.} 
\end{align}
The Mach number depends on $z$ through $r_g$, which is a function of the local temperature $T(r,z)$. 
 Here, if $r<r_{crit}(T)$, the sonic point is at $r_s$, where
 \begin{equation}
 r_s=\frac{r_g}{4}\left(1+\sqrt{1-r/r_{crit}}\right)
 \end{equation}
 When $r>r_{crit}(T)$, we set the flow Mach number to be unity (see Figure 4 of Waters \& Proga 2012).  We can thus define a critical gas temperature at a given radius $r$ which meets the sonic criterion, $r=r_{crit}=GM_*\mu/(8k_BT_{crit)}$ with $\mu=1.12\times10^{-24}$ g for fully ionized gas and $2.12\times10^{-24}$ g   for atomic gas.  Therefore, we have  T$_{crit} \sim 1.8\times10^4\ (1AU/r) $ K for a neutral atomic flow from a disk around a 1\ms\ star.  The vertical structure of the disk is such that the density rapidly falls with $z$ whereas the temperature (and hence $c_s$) rises more slowly as the heating  at the disk surface increases with $z$. At every $(r,z)$, we use the local temperature to evaluate $r_g, r_s$ and hence $\mathscr{M}$, and then find  $\dot{\Sigma}_{pe}$ for that radius using Eq.~\ref{perate}. Our procedure of finding the maximum $\dot{\Sigma}_{pe}$ with $z$ typically results in the flow being launched from a height where the temperature is slightly below this critical value (for $r>1$AU), at   $T_{gas}\sim 0.5-0.7\ $T$_{crit}$, with typical  flow Mach numbers $\sim 0.7-0.8$.
 
In order to achieve the goals of this paper, two key modifications have been made to the 1+1D disk structure model and the 1-D time-evolution model described above: dust evolution is now included, and gas temperature is determined differently. Details of each modification follow.

 \subsection{Dust evolution and 1-D disk models} 
The updated dust module of  our 1-D disk evolution model  now allows for various dust grain size distributions (32 logarithmically spaced size bins)\footnote{Throughout this work, dust grains are considered to be spherical and non-fractal in shape for simplicity.}. Dust and gas surface density distributions are evolved separately, with the gas component evolving as described by Eq.~\ref{sdot}. The time evolution of each dust component $\Sigma_{d}^i$ follows from the continuity equation where the rate of change of dust surface density is determined by the rate of diffusion. As the diffusion is related to the gradient in mass fraction, the evolution equation for  a dust particle of size $a^i$ is given by
\begin{equation}
\pd{\Sigma_d^i }{t} + \frac{1}{r}\pd{}{r}\left(r\Sigma_d^i u_d^i \right)  = \frac{1}{r}\pd{}{r}\left[
r\Sigma_g D_d^i \pd{}{r}\left(\frac{\Sigma_d^i}{\Sigma_g}\right)\right]
\label{eqsigmad}
\end{equation}
where $D_d^i = \nu/(1+St_i^2)$  is the dust diffusivity, $St_i \sim \pi a_i \rho_s/(2\Sigma_g)$ is the Stokes coupling parameter  and $u_d^i$ is the radial velocity due to gas drag and the radial pressure gradient.\footnote{We caution the reader that Eq.~\ref{eqsigmad} is no longer applicable when $\Sigma_d \sim \Sigma_g$, although we continue evolution in expectation of other planetesimal formation processes taking over at these epochs.} The dust drift velocity is given by
\begin{equation}
u_d^i = \frac{u_g(r)}{1+St_i^2} + \frac{1}{St_i+St_i^{-1}} \frac{1}{\rho_g \Omega} \pd{P}{r}
\label{vd}
\end{equation}
where $u_g$ is the gas radial velocity, $\rho_g$ the gas mass density and $\pd{P}{r}$ the radial pressure gradient. However, as discussed below, we renormalize the dust size 
distribution at a given $r$ after each time step, since coagulation and shattering are such rapid processes compared to radial advection.

Dust evolution is implemented in a number of incremental steps to isolate the effects of the various physical processes and understand their influence on disk evolution.  
\begin{enumerate}
\item Our simplest  prescription for the dust size distribution is an MRN power law, $n(a)\propto a^{-3.5}$ within a size range $a_{min}<a<a_{max}$, which is held constant in $r$ and $t$.  
We use the simple power law models as a reference with which to compare the more complicated effects of dust evolution.   Our fiducial reference model adopts the values $a_{min}=0.005\mu$m and $a_{max}=1$cm; the latter is an increase of over 4 magnitudes from the interstellar value. The dust opacity per H nucleus ($\sigma_H$) in this model, with a constant gas/dust mass ratio,  is lower by a factor of 100 from the interstellar value (for an MRN power law, $\sigma_H \propto 1/\sqrt{a_{min}a_{max}}$). Very small grains, including Polycyclic Aromatic Hydrocarbons (PAHs), heat gas via  the grain photoelectric effect and are significant in determining the mass loss due to FUV photoevaporation. Here, we typically assume that the PAHs are depleted in proportion to the very small grain abundance $(a_{min})$. Therefore, the PAH abundance in this case is also lower by a factor of 100 from the interstellar value. 
\item We next include the effect of collisional decoupling of dust grains with gas in the photoevaporative flow. Flow densities are typically such that larger grains collide too infrequently with the gas molecules to be carried along in the flow (see \S 4.1). 
 \item  The next model refinement is to consider a more realistic grain size evolution model using the fragmentation/coagulation equilibrium framework of BOD11.   BOD11 provide a 1-D prescription for the dust surface densities as a function of size ($\Sigma_d^i(a_i, r)$) that reproduces the more detailed results of their numerical simulations. $\Sigma_d^i(a_i, r)$
depends in this prescription on the total dust surface density, the gas surface density $\Sigma_g$, and $\alpha$.  The latter two determine relative collision speeds of grains that lead to fragmentation or coagulation.
 We thus have a consistent determination of the dust size distribution as a function of local disk quantities. Here again, PAHs are depleted in proportion to the calculated small grain abundances.  Eq.~\ref{eqsigmad} is used to determine the evolution of the dust surface densities. 
\item Finally, we include the effects of radial drift of dust grains with respect to the gas, using Equations ~\ref{eqsigmad} and \ref{vd}  to solve for dust evolution. 
\end{enumerate}

We summarize briefly the main features of the BOD11 models and their differences from the standard MRN size distribution of the ISM.  In these models, small dust grains coagulate and grow until they reach a larger fragmentation threshold where they shatter and replenish the small dust grain population. Although BOD11 describe several different regimes or size regions where various processes dominate, there are two critical sizes relevant to this work. One size ($a_{BT}$) dominates the cross-section and the other ($a_{frag}$) dominates the mass. For small grain sizes, Brownian motion determines the characteristic collision velocities for grains of size  $a_{min}<a<a_{BT}$.  Turbulence determines relative collision velocities for $a_{BT}<a<a_{frag}$ (also see Ormel \& Okzumi 2013). Relative velocities decrease with grain size for Brownian motion and increase with size for turbulent motion, and this defines the turn-over size $a_{BT}$ as
\begin{equation}
a_{BT}\approx 0.75 \left[\frac{\Sigma_g \mu m_p}{\alpha \sigma_{H_2}^{1/3} \rho_s}\right]^{3/10},
\label{abt}
\end{equation}
where $\mu m_p$ is the mean mass of the gas molecule, $\sigma_{H_2}$ is the cross-section of molecular hydrogen, and $\rho_s$ is the bulk mass density of the dust grains.   When particles reach a size $a_{frag}$, their collisional energies are large enough to shatter them, with $a_{frag}$ given by
\begin{equation}
a_{frag} = \frac{2 \Sigma_g}{\pi \alpha \rho_s} \frac{u_f^2}{c_s^2}, 
\label{afrag}
\end{equation}
where $u_f$ is the fragmentation threshold velocity.  We note that these sizes are defined for vertically integrated (with $z$) densities and hence it is $\Sigma_g$ that appears in the above expressions (BOD11). In the Brownian motion regime $a<a_{BT}$, the size-integrated dust mass of these small particles is proportional to $a^{3/2}$ (for grains not influenced by settling, see Table 3 of BOD11), and the contribution of grains of a given size to the total cross sectional area per H therefore increases with size ($\propto a^{1/2}$) for spherical grains. On the other hand, this contribution to the area  decreases with size ($\propto a^{-3/4}$) in the turbulent ($a>a_{BT}$) regime. Therefore, the total cross sectional area per H is dominated by grains of size $a_{BT}$. For grains not influenced by settling, the surface area is $\propto a_{BT}^{-3/4}$ (see Ormel \& Okuzumi 2013). There is thus a  significant reduction in the population of grains in the range $a_{min}<a<a_{BT}$ compared to the standard MRN power law. This is the main effect of the coagulation/fragmentation equilibrium model pertinent to gas heating which depends on the cross-sectional area per H ($\sigma_H$).  The deviations of the BOD11 distribution from the MRN power-law for grains in the size range $a_{BT}<a<a_{frag}$ are marginal, with a small increase in the number of the largest grains due to cratering effects. The collective mass in the grains is dominated by $a_{frag}$, which is dependent on local disk conditions and, in general, decreases with time. One key feature of this evolution is that as the disk evolves and the gas surface density ($\Sigma_g$) drops, the turbulent collisions of dust become more energetic, shattering dust and leading to a smaller $a_{frag}$. The transition from Brownian-motion-dominated collisions to turbulence-dominated collisions similarly shifts to smaller sizes, resulting in a smaller characteristic grain size $a_{BT}$ (Eqs.~\ref{abt} and \ref{afrag}).  For disk models where we include dust evolution, $\sigma_H$ at $t=0$ is smaller than the interstellar value (since $a_{max}$ is $\sim$ cm sizes compared to the sub-micron sizes in the ISM), and increases with time as the disk evolves toward lower surface densities. 

We assume that vertical size re-distribution in the disk occurs instantaneously, i.e. the grain size distribution is assumed independent of $z$  and the steady-state equilibrium grain size distribution is instantly attained.  Vertical mixing and grain growth are, in fact, rapid. Assuming collisions result in perfect sticking, the growth timescale $ \sim \Sigma_g/(\Omega \Sigma_d)$, or about $\lesssim 100$ orbital timescales. Vertical mixing is even faster (e.g., Dullemond \& Dominik 2005). Therefore, instantaneous re-distribution is a  reasonable simplifying approximation. At each instant of time, $\Sigma(r,t)$ and $\Sigma_d^i(r,t)$ are used to solve for the 1+1D structure to determine necessary disk parameters like the density and temperature structure as a function of $(r,z)$ and photoevaporation rate, and then evolved according to equations \ref{sdot} and \ref{eqsigmad}. At the end of the timestep, the updated surface densities and disk quantities are used to find the new dust size distributions at each $r$ using the prescriptions of BOD11. The total dust mass at every $r$ is then re-distributed appropriately into the various $\Sigma_d^i$ before advancing to the next timestep. The vertical and radial disk calculations are therefore decoupled, i.e., although we compute the disk vertical structure for each timestep, only the radial equations are evolved in time. We re-iterate that our disk evolution models are strictly 1-D, the 1+1D structure is only used to determine the photoevaporation rates input into Eq.~1.  

The separate treatment of gas and dust surface density evolution will yield the evolution of the gas/dust mass ratio in the disk as a function of radius and time.  There are three processes that can affect the gas/dust mass ratio in the disk: (i) radial drift of dust, (ii) photoevaporation, and (iii) dust coagulation to planetesimals. We only consider the first two processes. Radial drift can lead to a depletion of dust in the outer regions and a pile-up of dust at small radii and a change in the gas/dust ratio. This is calculated in our 2-fluid scheme because gas and dust (all sizes) are evolved separately.  We also determine the change in local gas/dust ratio by photoevaporation. Photoevaporation will carry small dust along with the gas in the flow, but grains larger than some minimum size will be decoupled from the gas. This minimum size $a_{pe}$ is determined from a force balance between the drag force acting on a dust grain due to collisions with gas, and the gravitational force that acts to cause the dust grain to settle in the vertical direction (see \S 4.2). Particles larger than $a_{pe}$ will not be carried away with the flow and are left behind in the disk.  This selective removal of small dust with the gas and  the retention of large dust grains lowers the gas/dust mass ratio in the disk. Lastly, dust coagulation changes the gas/dust mass ratio if a significant fraction of the dust has evolved into large objects which lock up most of the mass of solids. We only consider dust grains that are smaller than 1cm; grains larger than $\sim 1$cm are not collisionally well-coupled to the gas and because of their small numbers also present very little absorption to stellar optical photons, rendering them unimportant for gas-dust collisional heating and determination of disk vertical structure.  At sizes  $a\gtrsim1$cm, other processes not considered here such as trapping in vortices and turbulent concentration of particles are expected to lead to the formation of planetesimals (e.g., Chiang \& Youdin 2010).  However, these processes and the barriers to growth beyond $\sim 1$cm are not well understood (e.g., Testi et al. 2014). Hence, we do not consider dust coagulation from cm sizes to planetesimals as a source for draining the dust mass in the disk.

\subsection{Gas Temperature and Simplified Chemistry}
GDH09 used a look-up table approach to solve for the gas temperature to eliminate the need for solving a chemical network. The gas temperature  was calculated as a function of local variables (including density and radiation field) with a $k-$nearest neighbor ($k$-NN) method using a large database of results from detailed thermochemical models (Gorti \& Hollenbach, 2004, 2008). With the addition of dust properties, the dimensionality of the parameter space increases to a point where the $k-$NN approach rapidly becomes impractical. We have changed this method of calculating gas temperature as it does not permit exploration of a range of stellar and disk conditions and dust properties.  We now retain most essential aspects of our more detailed thermochemical models (GH08), such as the heating mechanisms, calculation of local attenuation in the disk for FUV heating and dust radiative transfer, and the absorption column determination for X-ray ionization and heating rates. However, we use a new simplified approach for gas cooling and disk chemistry. We thus drastically reduce the computational overhead in solving for the chemical network, which is needed if we are to include dust evolution as described in \S 3.1. The loss in accuracy is estimated to be $\sim 10-25$\% compared to the thermochemical models with full chemistry, and is comparable to the previous methods. Given the exploratory nature of the intended parameter survey and the inherent uncertainties in many of the other input parameters (e.g., reaction rates, microphysics), we find this to be a feasible approach. Further, our objective is to look for trends in evolution when various parameters are varied, and the uniform new method for determining the gas temperature is adequate for this purpose.  

The new simplified chemistry can be summarized as follows. We calculate the H/H$_2$ transition in the disk by equating H$_2$ formation  (due to 3-body reactions, H$^{-}$ and on grain surfaces) with its destruction (due to photodissociation by FUV\footnote{self-shielding is considered as well} and X-ray photons and by reactions with atomic oxygen to form OH). We determine the ionization fraction ($x_e$) due to X-rays based on the local X-ray flux and use a simple parametric relation for the electron abundance due to FUV photons based on results from our thermochemical models (GH04, GH08). We  add these together to obtain a total electron abundance. We ignore cosmic rays, our region of interest is the photoevaporating disk surface where cosmic ray ionization is not important. X-ray ionization is determined as a function of local attenuation column (see GH04 for details) and the ionization fraction is determined from a balance between ionization and recombination (Shull \& van Steenberg 1985, Glassgold et al. 1997).  The electron abundance due to FUV ionization is approximated as  
$x_e = 10^{-3.73}(G/n)^{0.42} {\rm \ for\ } G/n < 10^{-4} ;
x_e = 10^{-3.7} {\rm \ for\ } G/n > 10^{-2.5} ; {\rm and\ } 
x_e = 10^{-1.2}(G/n)^{1.05} $ for all intermediate values of $G/n$. 
$G$ is the local attenuated FUV flux in Habing units and $n$ is the gas number density in cm$^{-3}$. The electron abundance due to FUV alone is restricted to  a maximum value of $10^{-3.7} $, which corresponds to a situation when all the gas phase carbon in the disk is ionized. The C/C$^{+}$ photoionization front is determined by balancing FUV photoionization with recombination. We then assume that when FUV shielding enables C$^+$ to recombine (i.e., at the C ionization front), the carbon transitions to CO. We ignore any possible intermediate zone, typically small in depth, where carbon may be mostly atomic.  Above the C$^+$ ionization front, Fe and S are assumed to be ionized. The ionization of neon and argon is by X-rays and these are determined analytically (e.g., as in Glassgold et al. 2007). Since the recombination rates depend on $x_e$, we iterate this process to obtain a consistent solution at a given $z,r$ and $t$.

\begin{figure}[f]
\centering
\plottwo{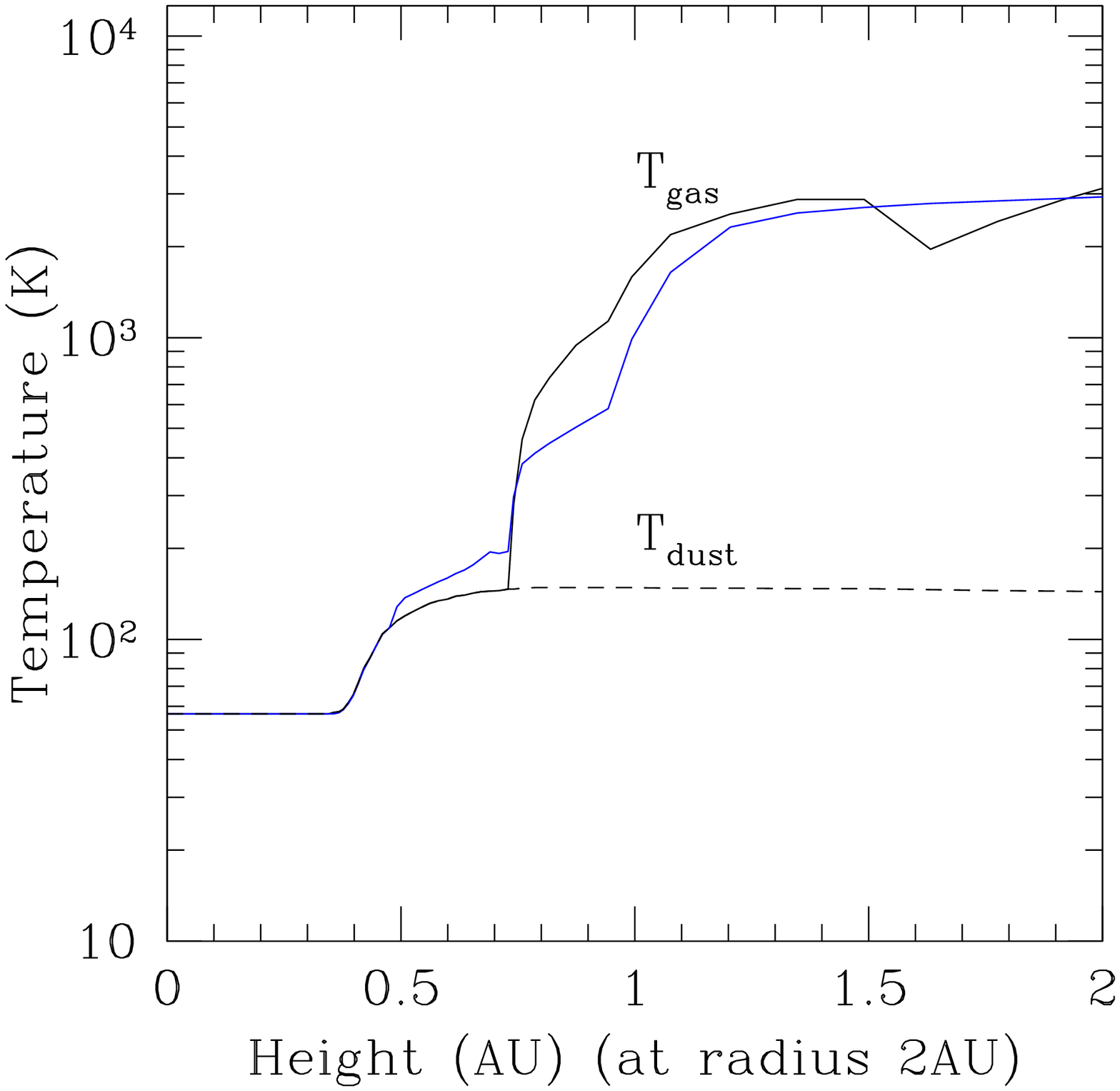}{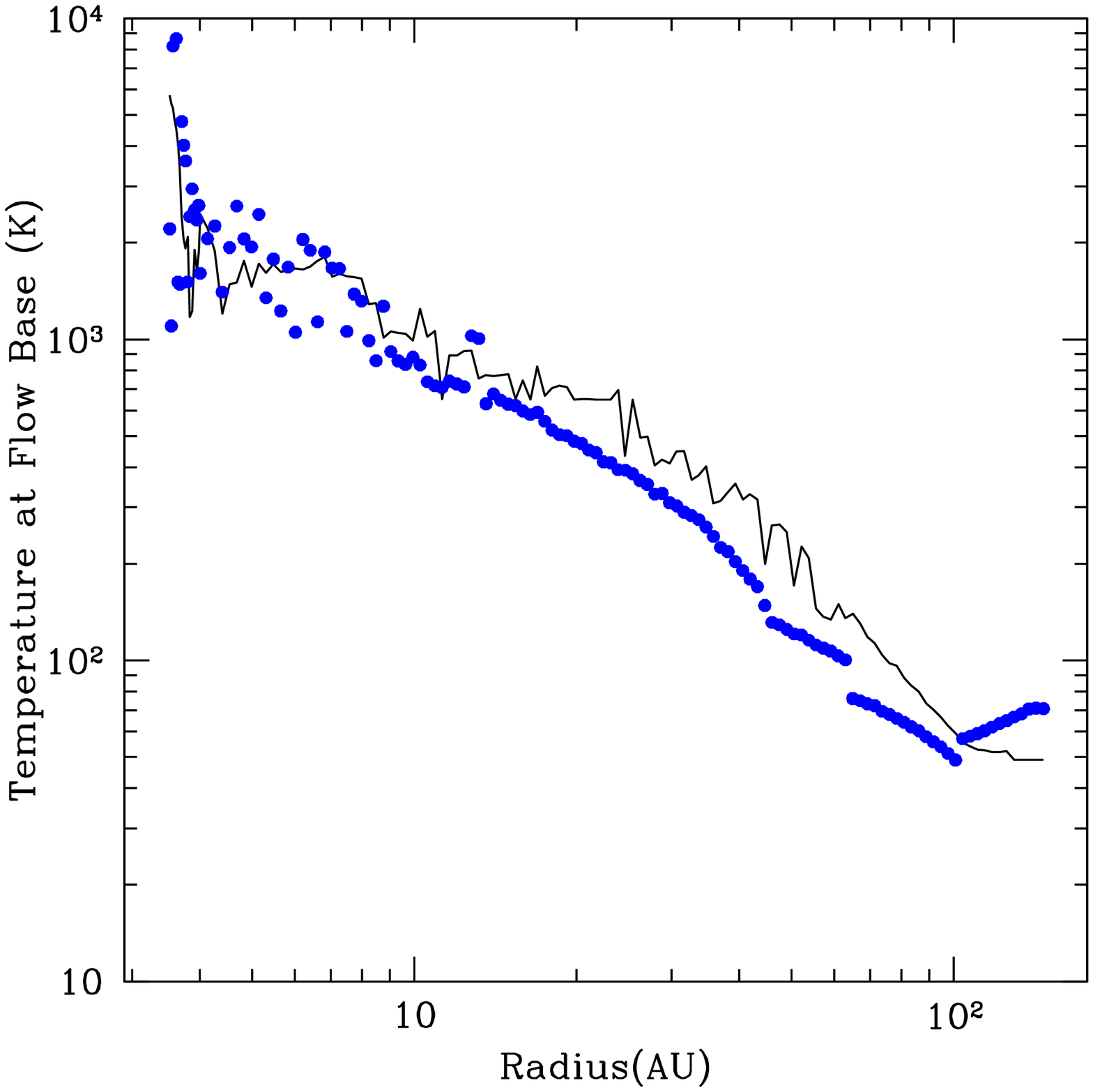}
\caption {\it The gas temperature at a radius of 2 AU as a function of height using the simplified model (solid black line) approximation as compared to a full thermochemical disk calculation (blue line) for a model disk. The dashed line indicates the dust temperature at that height. Also shown in the right panel are the gas temperatures at the layer ($z(r)$) where the photoevaporative flow originates for the simple (black line) model used here and the full thermochemical models (blue circles). }
\label{tvert}
\end{figure}

We calculate the gas temperature at a given spatial location by solving for this reduced chemical network with thermal balance between heating and gas cooling in the 1+1D disk structure determination at each time epoch. We consider thermal energy exchange due to collisions with dust, heating by FUV photons and X-rays, and chemical heating due to pumping and dissociation of H$_2$ and ionization of carbon.  Cooling is due to [OI], [CII], [SiII], [SiI], [SI], [SII], [FeI], [FeII], [NeII], [NeIII], [ArII] and [ArIII] fine structure lines, [OI], [SiI], [SiII], [SI], [SII], [FeI], [FeII] optical and near-infrared lines, Lyman $\alpha$, H and He recombinations, and H$_2$ and CO rotational and vibrational lines.   Molecular cooling is implemented with the help of analytic cooling functions (e.g., Hollenbach \&McKee 1979, 1989).  We find that this approximated gas temperature reproduces the results of the more detailed chemical models reasonably well (see Figure~\ref{tvert}a); differences between the approximate and full model calculations are typically small at the location of the base of the photoevaporation layer and the derived photoevaporation rates (where $\dot{\Sigma}_{pe} \propto \sqrt{T}$)  are similar. Figure $\ref{tvert}$b shows the radial temperature profile at the base of the photoevaporative layer for a model disk using a full thermochemical model and the simplified procedure described above. 

\section{Results and Discussion}
Dust evolution modifies the spatial and temporal distribution of dust grains in the disk, affecting disk opacity and the penetration of stellar optical and FUV radiation through the disk, and hence, disk thermal and chemical structure. To understand which dust evolutionary process has the most significant effect on disk mass and lifetimes, we consider these in progressive refinements as described in \S 3.1. (Readers interested mainly in the final model results may skip to \S 4.4.)
We first describe photoevaporation in a reference model with median dust properties and no dust evolution. While the assumption of a dust distribution static with time and radius is clearly unrealistic, the reference model nevertheless provides a useful average evolution scenario with which to compare the other models to be described below.  This discussion is followed by the results of parameter studies where we assume a fixed slope and minimum size to the dust grain size distribution and vary disk flaring angles, maximum dust grain size and the gas/dust mass ratio.   We then calculate the critical dust grain size ($a_{pe}(r,t)$) that is collisionally coupled to the photoevaporative flow and retain dust particles with sizes larger than $a_{pe}$ in the disk.  We next relax the assumption of a constant size distribution and include the dust coagulation/fragmentation model of BOD11. Finally, we include the effects of radial drift of dust grains. We then explore the effects of variations in PAH abundance, X-ray luminosity and the viscous parameter $\alpha$ on disk evolution and lifetimes. 

\subsection{Reference Model with No Dust Evolution}
\begin{figure}
\centering
\plotone{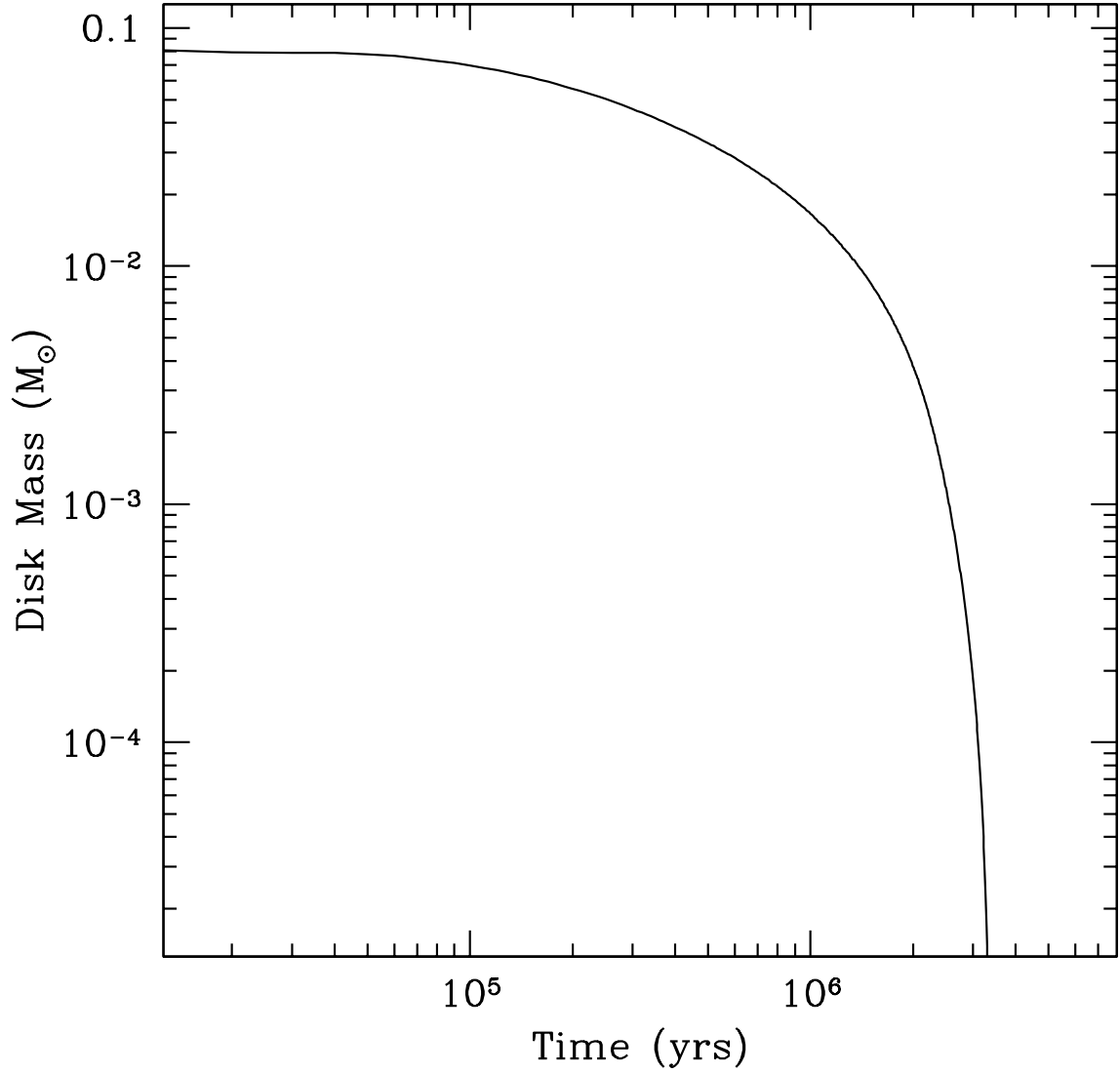}
\caption{\it  The disk mass as a function of time for the reference  model with chromospheric XEFUV and accretion-generated FUV (``Gap Model" of Fig.~\ref{overview}). This disk has a fixed, non-evolving, dust size distribution (in radius and time) with an MRN power law and size $0.005\mu$m $<a<$1 cm. }
\label{defcase}
\end{figure}

Our reference model is that of a disk of initial mass $0.1$\ms\  around a 1\ms\ star, with UV and X-ray luminosities and all other relevant input parameters as given in Table 1.  Our default choice of the viscous parameter hereon is $\alpha=0.01$.  Disk evolution is essentially as described in GDH09, but with the temperature determined as described in \S 3.2. In this model, dust does not evolve --- grain size distribution is the same at all radii, dust is assumed to remain coupled with the gas even in the flow, and there is no radial drift of dust with respect to gas. There is thus no change in the gas/dust ratio in this model.  Dust grain sizes assumed here are $0.005\mu$m$-1$cm, in the range inferred from SED modeling (e.g. Williams \& Cieza 2011).  Figure~\ref{defcase} shows the disk mass as function of time. The disk survives for 3.5 Myrs.\footnote{Numerically obtained disk lifetimes hereafter are defined by the time when $\Sigma_g(r,t)<10^{-20}$ g cm$^{-2}$ at all radii, or $M_{disk}\lesssim10^{-14}$\ms.  We note that a different definition (e.g., $\Sigma_g(r,t)<10^{-25}$ g cm$^{-2}$)  does not change the results of the study.}

We consider two limiting cases with high and low chromospheric XEFUV (X-rays, FUV and EUV) luminosities---the level of chromospheric flux determines whether or not the disk evolves by opening gaps. We first qualitatively discuss disk evolution with viscosity and photoevaporation, and then present detailed, numerical models of these two cases. Typically, in the case of relatively high chromospheric XEFUV, the disk evolves in time so that at some inner radius, the net incoming mass flux due to accretion drops to zero because of significant mass loss due to photoevaporation beyond this radius and a gap opens (e.g., GDH09). Once the inner and outer disk are decoupled at the gap, the inner material viscously accretes onto the star and a hole is formed. The disk further evolves by irradiation of the outer rim due to direct photoevaporation (Alexander et al. 2006). However, in the case of very low XEFUV chromospheric luminosity, gaps and holes form so late in the disk evolution that virtually no disk material remains. In this case, although the initial accretion-generated FUV (or possibly XEFUV) is high, the declining rate of accretion rapidly lowers the production of FUV, and this feedback prevents the early onset of a gap. The gap only forms when the very weak (but constant) chromospheric XEFUV can sustain it.

Disk dispersal timescales can be estimated by comparing the viscous mass accretion rate $\dot{M}_{acc}$ to the photoevaporation rate $\dot{M}_{pe}$  (e.g., Clarke et al. 2001). In order to qualitatively understand disk dispersal, we consider analytical solutions  for a purely viscous disk (Lynden-Bell \& Pringle 1974), under the simplifying assumption that the viscosity $\nu \propto r$. (The calculated radial temperature profile in the midplane of the disk yields a slightly steeper dependence of $\nu$ with $r$ for constant $\alpha$).   The accretion rate is nearly constant with $r$ and is given by
\begin{equation}
\dot{M}_{acc} = \frac{M_d(0)}{2 t_s}\left(1+\frac{t}{t_s}\right)^{-3/2},
\label{sigmav}
\end{equation}
where $t_s$ is the viscous timescale at a radius $R_1$ outside of which there is $1/e$ of the initial disk mass $M_d(0)$ and is given by $t_s = R_1^2/(3\nu(R_1))$ (e.g., Hartmann et al. 1998). For a 100AU disk with $\Sigma \propto r^{-1}$, for example, $R_1\sim 60$AU. Photoevaporative timescales, $t_{pe0}=M_d(0)/\dot{M}_{pe}$, are typically much larger than $t_s$. Note that $\dot{M}_{acc}$ decreases with time, whereas $\dot{M}_{pe}$ may remain relatively constant. 
The dispersal timescale $t_{dispersal}$ when $\dot{M}_{acc}=\dot{M}_{pe}$  is therefore, 
\begin{equation}
t_{dispersal} \approx 0.63  \left(\frac{M_d(0)}{\dot{M}_{pe}}\right)^{2/3} t_s^{1/3} 
\approx 0.63 \left(\frac{t_{pe0}}{t_s}\right)^{2/3} t_s.
\label{tgap}
\end{equation}
The disk thus disperses on a certain number of viscous timescales, determined by the ratio $t_{pe0}/t_s$. For a disk with $M_d(0)=0.1$\ms, $\dot{M}_{pe}\sim 10^{-9}$\mpy,  and  $\alpha=0.01$, $t_s\sim 4.5\times10^5$ years, $t_{pe0}=10^{8}$ years and
$t_{dispersal}\sim 10^{7}$ years. If $\dot{M}_{pe}$ is constant, $t_{dispersal}$ is more sensitive to $t_{pe0}$ than to $t_s$, and photoevaporation is important in setting
$t_{dispersal}$.  Moreover, from Eq.~\ref{tgap}, $t_{dispersal}\propto \alpha^{-1/3}$ for an $\dot{M}_{pe}$ independent of $\alpha$. If $\dot{M}_{pe}$ changes with time or depends on $\alpha$, this proportionality changes, as we show later. 

For a disk with high chromospheric XEFUV, dispersal is rapid after the hole forms.  Early in the disk's evolution, accretion-generated FUV drives significant mass loss from the outer disk and along with the chromospheric XEFUV creates a gap in the inner disk (see GDH09). After quick viscous draining of the inner disk, direct illumination of the rim by the high chromospheric flux of XEFUV photons lead to vigorous photoevaporation  and the disk is eroded outwards to disperse from the ``inside-out".  The disk disperses within a few additional viscous timescales after gap opening.  This scenario---in the case of pure chromospheric EUV and X-rays ---corresponds to the models of Alexander et al (2006) and Owen et al. (2010).   The addition of accretion-generated FUV and chromospheric FUV increases $\dot{M}_{pe}$ over the disk by strong photoevaporation of the outer disk and its truncation, and by enhancing  inside-out evaporation through a further increase in the mass loss at the rim. 

The other limiting case is where the high energy photons are mainly supplied by accretion onto the stellar surface, and the chromospheric component is absent or very weak. Here, a gap cannot be sustained by the accretion-generated XEFUV, and cannot form if there is no chromospheric component to the XEFUV. For purely accretion-generated XEFUV flux, the disk loses mass steadily both by viscous accretion onto the star and photoevaporation of the outer ($> 1$ AU) regions. Continued accretion generates a steady but decreasing supply of high energy photons that drive photoevaporation until the disk is dispersed from the Òoutside-inÓ.

In either case, the disk lifetime is determined by the rate at which the disk loses mass. If $\dot{M}_{pe}$ is not too different for the high and low chromospheric XEFUV cases, similar disk lifetimes are to be expected although disk evolution is qualitatively quite different. As we show below, more exact numerical modeling confirms these expectations. 

\begin{figure}[f]
\centering
\plottwo{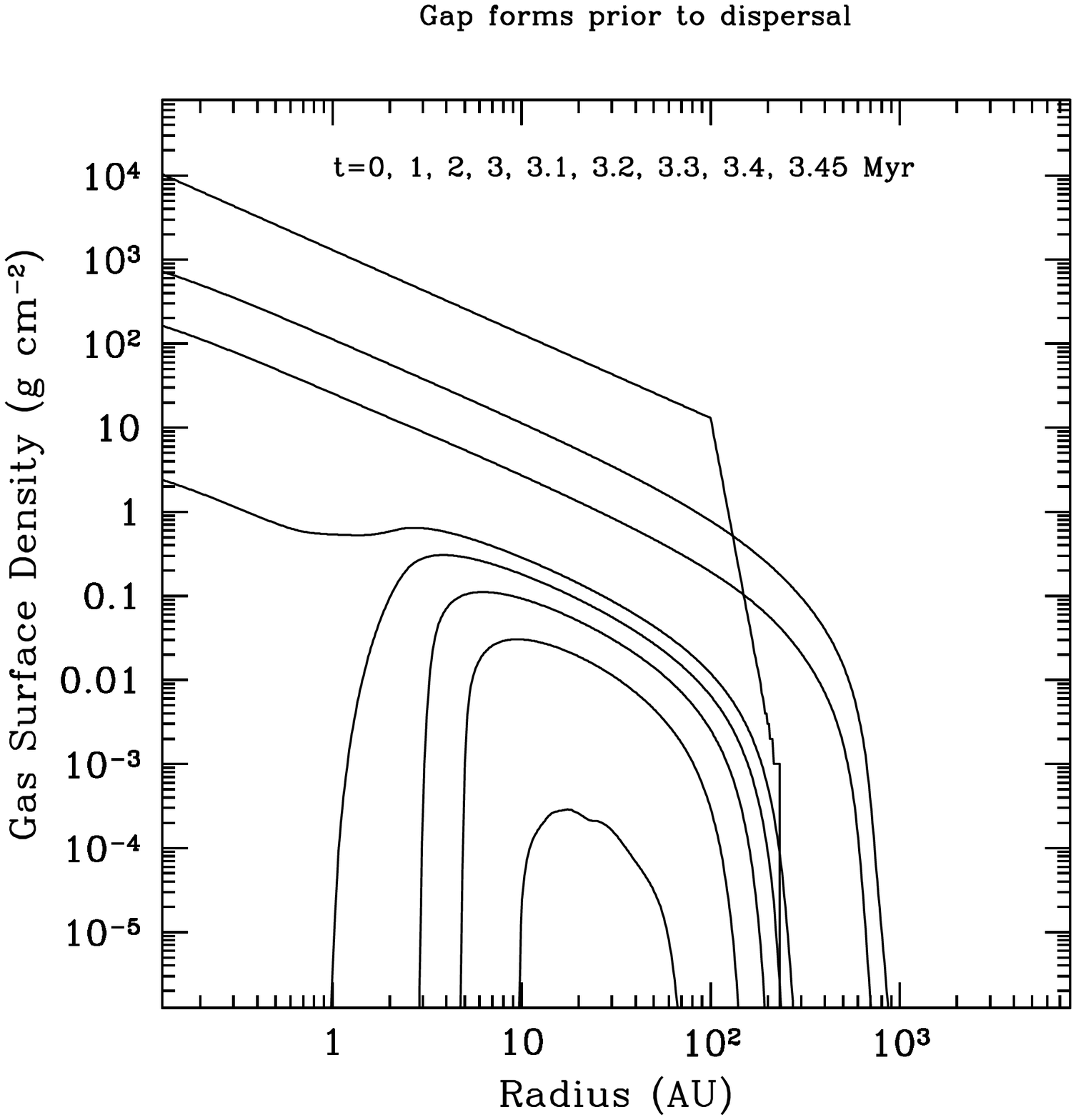}{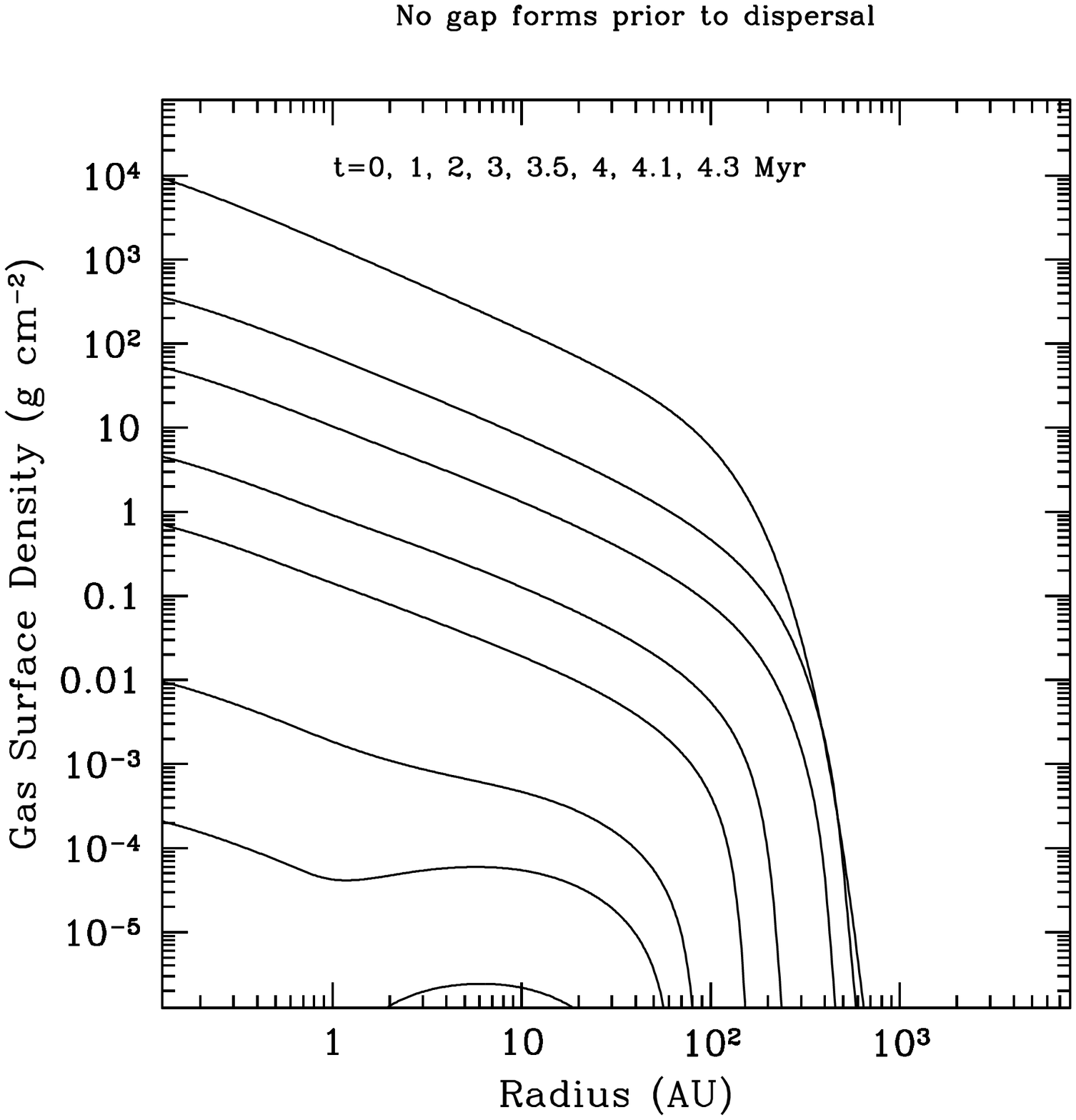}
\caption{\it Evolution of two disk models with and without gap formation is shown.
The panel on the left shows the surface density evolution of a model disk with a high chromospheric luminosity (FUV, EUV and Xray luminosities of $2\times10^{30}$ erg s$^{-1}$ where a gap opens at $\sim 1$AU at $3.1$Myr before the disk is dispersed.  After gap opening, XEFUV photons irradiate the rim of the inner hole. The disk disperses in 3.5 Myrs. In the panel on the right, a low chromospheric component is assumed  ($10^{28}$ erg s$^{-1}$), so that an inner hole cannot be sustained without accretion and the accompanying flux of high energy photons.  In this case,  the disk evolves to shrink radially with time and disperses in 4.3  Myrs.}
\label{overview}
\end{figure}

As discussed in \S 3, accretion and chromospheric activity are both responsible for the production of XEFUV photons in young stars.  We now present two detailed numerical disk models which take into account stellar accretion and chromospheric activity as a source of EUV, FUV and X-rays.  Figure~\ref{overview}  depicts the results of two model runs: (i) a ``Gap" model which has a time-dependent accretion generated FUV, and a constant chromospheric emission of FUV, EUV and X-rays with a luminosity of $2\times 10^{30}$ erg s$^{-1}$ each, and (ii) a  ``No Gap" model that includes a time-dependent accretion component for FUV, EUV and X-rays but where the chromospheric emission for all three is low (at $10^{28}$ erg s$^{-1}$).  Although the contribution of accretion to EUV and X-rays is less certain (e.g., G\"{u}del \& Telleschi 2007) than for FUV (e.g., Calvet \& Gullbring 1998),  we have assumed in this one particular case that accretion shocks produce a predominantly soft ($T_X=5$MK) X-ray flux and that this component saturates at $10^{-3}$L$_{bol}\sim 4\times 10^{30}$ erg s$^{-1}$ (e.g., G\"{u}del \& Naze 2009). EUV is assumed to be at the same level as the X-rays, and FUV is treated as before.  These assumptions ensure that photoevaporation rates  for the two models are not too different, and therefore that  in both cases the surface density evolution is similar through most of the disk lifetime to allow comparisons between the two models. 
 
The surface density evolution of the two model disks is qualitatively similar with an initial $\sim 1$Myr viscous expansion epoch where photoevaporation causes the outer radius to decrease with time, right until the gap opening epoch at $\sim 3$ Myrs after which it differs substantially (Fig.~\ref{overview}). In the ``Gap" model, a gap opens at 1AU at this epoch and the disk disperses rapidly in $\lesssim 0.5$ Myrs thereafter. In the ``No Gap" model, the disk continues to shrink radially in extent as it dissipates and survives slightly longer, $\sim4.3$ Myrs. The slight increase in lifetime is due to the declining flux of accretion-generated XEFUV, which makes $\dot{M}_{pe}$ for the ``No Gap" model lower than the ``Gap" Model at $t\gtrsim3$  Myrs. For these models which begin with a $0.1$\ms\ disk, $t_s=4.5\times10^6$ years and the time averaged $\dot{M}_{pe}\sim 5\times 10^{-9}$\mpy, which according to Eq.~\ref{tgap} yields disk lifetimes of $\sim 3.6\times10^6$ years and agrees with the numerical result.  As we might expect from Eq.~\ref{tgap}, because $t_{pe0}$ and $t_s$ for both models are similar, disk lifetimes  are not very different. In both cases, disk mass lost by photoevaporation is similar to the disk mass lost by accretion onto the star. 

 We hereafter always assume that the radiation field is as described in the ``Gap Model" above with accretion producing only FUV and a chromospheric XEFUV radiation field. This is also our fiducial reference model from hereon. In the following we describe the effects of varying various dust-related properties on the spatial and temporal evolution of the surface density (and mass) in the disk. 

\subsection{Parameter Studies with a Non-evolving Dust Size Distribution}
\paragraph{ Flaring angle} 
The disk flaring angle ($\alpha_{fl} \sim d(H(r)/r)/d \ln r$) is the curvature of the disk scale height($H(r)$) as a function of radius($r$), and determines how the disk surface is shaped. 
The optical depth of unity layer which separates the optically thick midplane from the superheated optically thin surface is higher in $z$ for more flared disks which are vertically more extended. Higher flaring angles imply higher solid angles subtended by the disk and hence greater interception of stellar photons.   This is likely to result in an enhancement of the disk photoevaporation rate and it therefore might be expected that disk lifetimes would decrease as the flaring angle increases. We computed a series of models with constant flaring (at all $r$ and $t$).  Indeed, we find that less flared disks do live longer than more flared disks, but even large changes in $\alpha_{fl}$ from 0.01 to 0.3 only result in a change in disk lifetimes from 4.7 Myrs to 3.1 Myrs. The disk lifetime is thus not a very sensitive function of the flaring angle. We therefore  do not compute the flaring angle as a function of radius at every epoch in the parameter survey below. Determining $\alpha_{fl}$ from the disk curvature is also inherently unstable and could lead to numerical issues. We instead {\em average} the locally determined $\alpha_{fl}$ over the disk radius at each timestep, and use this averaged value at all radii for the next timestep. We adopt a floor value of 0.01 for  $\alpha_{fl}$, below this the disk is effectively self-shadowed and cannot be solved within our 1+1D framework. 

\paragraph{Dust cross-sectional area: Size distribution and gas-to-dust mass ratio} 
The relevant physical quantity for gas thermal balance is the collective surface area presented by the dust grains, as we show here. The grain cross-section per H atom, $\sigma_H$, determines the cooling (or heating) of gas by collisions with dust grains, FUV heating through the coupling of the PAH abundance with $\sigma_H$, penetration depth of FUV photons, gas-grain drag and the rate of formation of H$_2$ on grain surfaces. 

\begin{figure}
\centering
\plotone{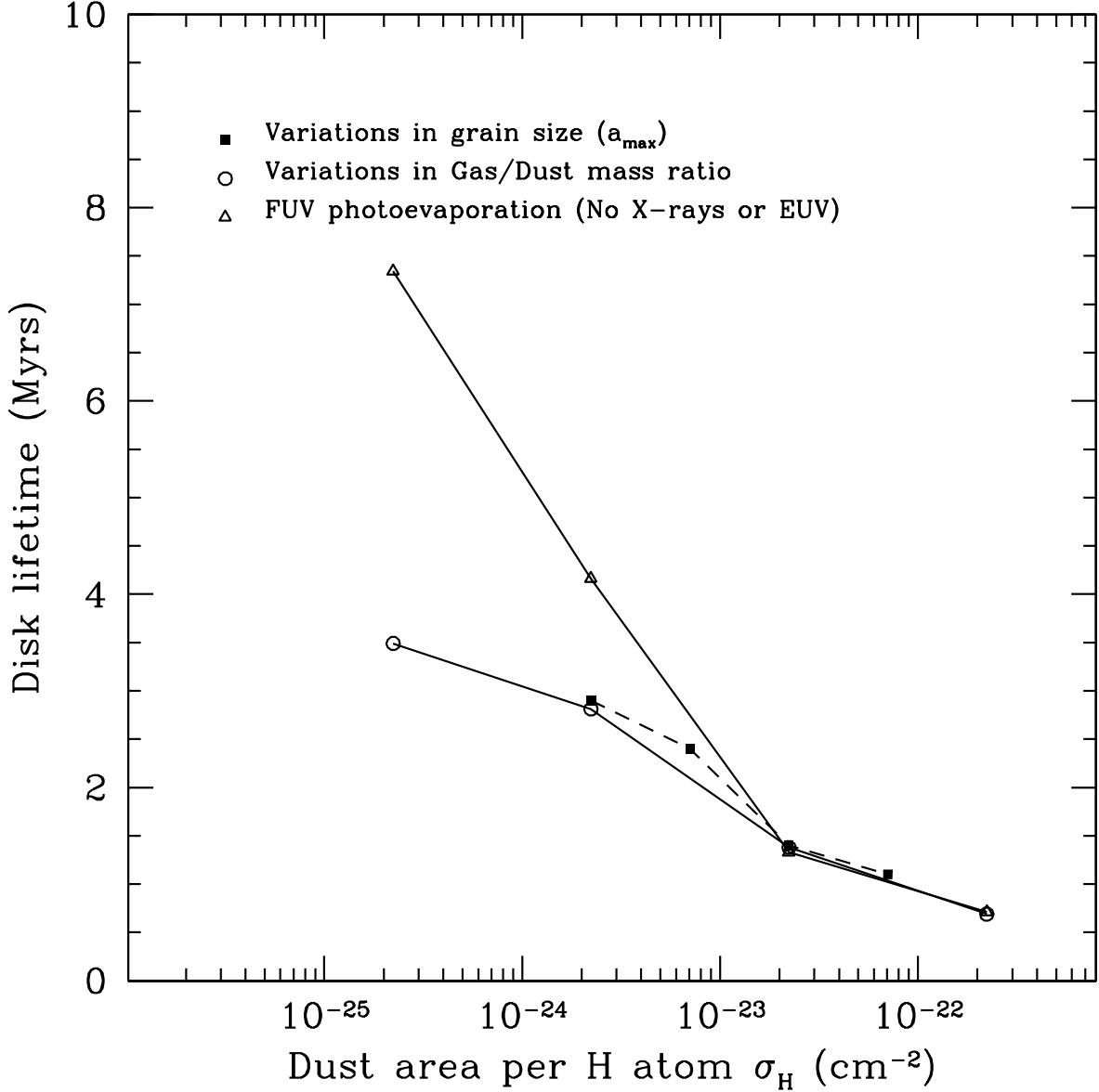}
\caption{\it The effects of the dust cross sectional area $\sigma _H$ (per H atom) on disk lifetimes are shown. The open circles show a change in the gas/dust mass ratio ($\eta$) with a constant maximum grain size ($a_{max}$=1cm), and in this case $\sigma_H=2.3\times10^{-24}(100/\eta)$ cm$^{-2}$ per H. The four cases shown represent gas/dust mass ratios of $10^3,10^2, 10$ and $1$. The filled squares are models where the gas to dust mass ratio is held constant at 100 but the maximum grain size in the distribution is varied. Here $\sigma_H=5.0\times10^{-22}\sqrt{0.2\mu m/a_{max}}$ cm$^{-2}$ per H. The four models are for values of $a_{max}=10\mu$m, 100$\mu$m, 1000$\mu$m and 1cm. The minimum grain size is held constant at $0.005\mu$m. The triangles show models with different $\eta$ but with photoevaporation only due to FUV photons (no X-rays or EUV). For reference, $\sigma_H=5.0\times10^{-22}$ cm$^{-2}$ per H for dust in the ISM. }
\label{sigmah}
\end{figure}

Dust grains are assumed to follow an MRN size distribution in these simple models, where the cross-sectional area is dominated by small grains and the mass is dominated by large grains;  the number density $n(a) \sim a^{-3.5}$, for $a_{min}<a<a_{max}$ (e.g. Dohnanyi 1969, Weidenschilling 1977). $a_{min}$ is kept fixed at 0.005$\mu$m.  For an MRN power law and fixed $a_{min}$, $\sigma _H$ can be increased either by decreasing the maximum size ($a_{max}$) of the dust grains which dominate the mass, or equivalently by decreasing the gas/dust mass ratio($\eta$).  Since dust mass is conserved, $\sigma_H \propto (a_{min}a_{max})^{-0.5}$, and decreases as $a_{max}$ increases. The change in disk lifetime is shown in Figure~\ref{sigmah} for a set of models with different $a_{max}$; as we explain below, disk lifetimes get shorter with increasing $\sigma_H$(decreasing $a_{max}$). In the next set of runs, the gas to dust mass ratio ($\eta$) in the disk is varied over 4 orders of magnitude. The resulting change in disk lifetime is again shown in Figure~\ref{sigmah} (open circles).  The two sets of models are in agreement for equal values of $\sigma_H$, indicating that $\sigma_H$ is the relevant parameter as expected. 

The near invariance of disk lifetime with $\sigma_H$ seen in Fig.~\ref{sigmah} can be partly attributed to gas heating and cooling mechanisms.  For large $\sigma_H$, heating is mainly by FUV photons and cooling is due to collisions with dust; as they are both linearly proportional to $\sigma_H$ this results in a disk density/temperature structure that varies very little with $\sigma_H$, resulting in similar disk lifetimes. For small $\sigma_H$, heating and cooling by dust collisions and heating by FUV photons decreases because of the lower abundance of small grains, making X-rays more significant for heating. Cooling is mainly due to line emission by [OI] and other species.  Since these are independent of $\sigma_H$, the disk lifetime due to photoevaporation remains fairly constant in the low $\sigma_H$ range.

However, even in the case of no X-rays and pure FUV photoevaporation, low $\sigma_H$ and lower heating rates do not result in a significant increase in disk lifetimes (see Fig.~\ref{sigmah}).   In models with FUV heating alone, disk lifetimes at low $\sigma_H$ change only by a factor of $\sim 7$  for a factor of 100 change in $\sigma_H$ and thereby the FUV heating rate in unattenuated surface regions. The increased heating 
is partially offset by the fact that the increased $\sigma_H$ also increases dust attenuation 
of the FUV and therefore reduces the penetration of FUV photons. As a result, the density 
at the base of the photoevaporative flow (where $T \sim T_{crit}$) is less than in the case of the same FUV flux but without the extra attenuation included.

\subsection{Photoevaporative reduction of gas/dust mass ratio}
Small dust  that is collisionally coupled to the gas is carried along with the photoevaporative wind, but larger dust particles may decouple from the flow and be left behind in the disk. The critical dust grain size $a_{pe}$ that is carried with the flow can be found by equating the drag force due to collisions with the downward gravitational force experienced by the dust grain. For typical disk and flow conditions, the mean free path for gas particle collisions is much larger than the size of the dust particle, placing the drag force in the Epstein regime.  $a_{pe}$ can thus be estimated; 
\begin{equation}
{ {8 \sqrt{2\pi} }\over{3}} \rho_{gas} a_{pe}^2 c_s \Delta v = ({{4\pi}\over{3}}\rho_{grain} a_{pe}^3) \Omega^2 z \hspace{0.5cm} {\rm or,} \hspace{0.3cm} a_{pe} \sim \frac{2\sqrt{2}}{\sqrt{\pi}} {{\rho_{gas}}\over{\rho_{grain}}} {{c_s^2}\over{\Omega^2 z}}
\label{eqape}
\end{equation}
where $\rho_{gas}$ is the gas mass density,  $\rho_{grain}$ is the bulk density of the dust grain material and $\Omega$ is the Keplerian angular frequency at radius $r$. The height $z$ is the location of the base of the wind. Wind speeds approach the thermal speed $c_s$ (see \S 2), and the relative velocity between gas and dust ($\Delta v$) must therefore be of this order, which leads us to the approximate expression for $a_{pe}$ in the above equation. At 200 AU, for example, we find from models that $\rho_{gas} \sim 2\times10^{-19}\text{g cm}^{-3}$ and that $T_{gas}\sim T_{crit}$, which gives $a_{pe}\sim 2\mu$m.  Larger particles will have a net speed downward into the disk and  smaller particles are carried along upward in the flow, changing the gas/dust mass ratio in the disk.
For the above case at 200 AU, since $a_{frag} \gg 2\mu$m and most of the mass resides in these larger grains, most of the dust remains behind in the disk and the photoevaporative flow is essentially pure gas at this radius.
  The assumption $\Delta v \sim c_s$ gives an upper limit on the gas/dust mass ratio.  If the flow velocity were substantially sub-thermal, then $a_{pe}$ would be correspondingly smaller than that estimated from Eq.~\ref{eqape} and more dust  may be retained.

\begin{figure}
\centering
\plottwo{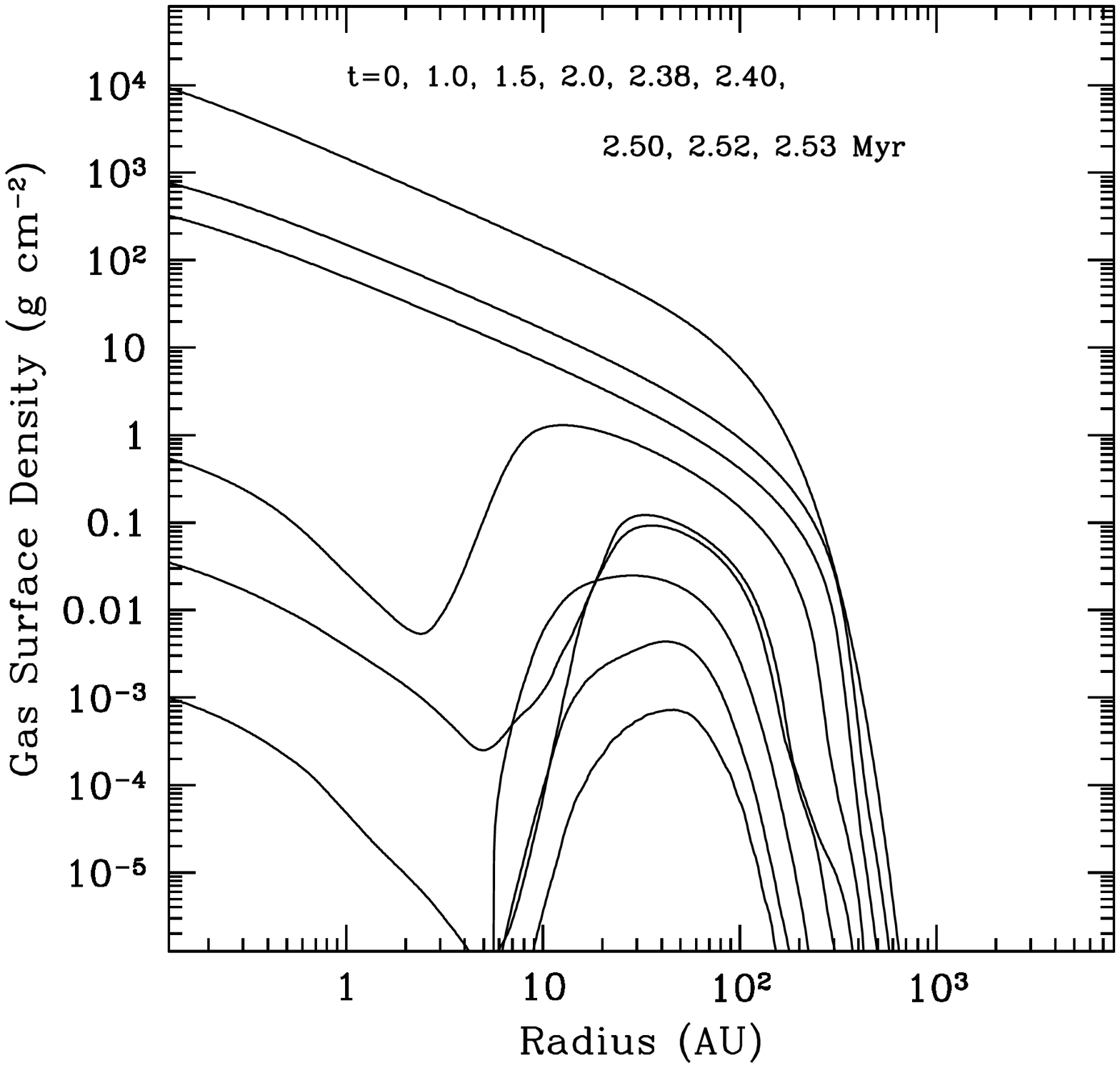}{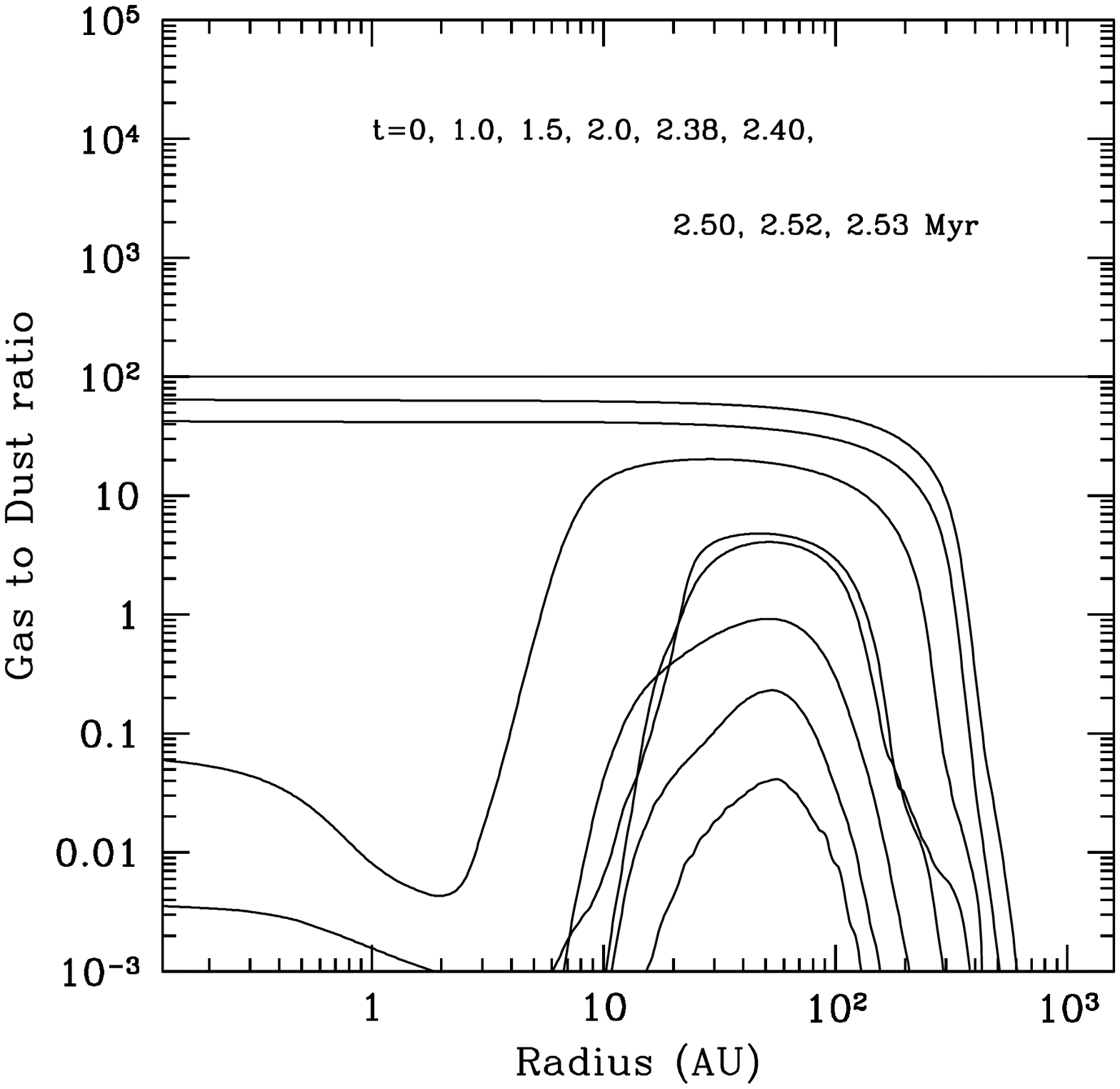}
\caption{\it  Disk evolution with a constant MRN dust size distribution and loss of small grains with photoevaporation is shown. The gas surface density as a function of radius for several different epochs is shown in the left panel; the right panel shows the corresponding gas/dust mass ratio in the disk. For $t\lesssim2$ Myrs, the disk accretes viscously onto the star while the outer disk loses mass by photoevaporation. Vigorous flows in the outer disk carry very little dust and the gas/dust ratio decreases at these radii. Viscosity simultaneously redistributes disk material to result in a reduction in the gas/dust mass ratio across the entire disk. When a gap opens in the inner disk at $\sim 2$ Myrs, the inner and outer disk are decoupled and almost all the dust mass beyond the gap edge is retained while the disk is dispersed. The gas/dust ratio over the entire disk thus rapidly decreases with time after 2 Myrs.  }
\label{ape}
\end{figure}

 We include this criterion (Eq.~\ref{eqape}) in the disk model to follow the change in gas/dust mass ratio while keeping the dust size distribution fixed according to an MRN power law (Table 1).   Starting from the initial ($t=0$) surface densities for grains of different sizes, smaller grains are removed by photoevaporation while larger grains are retained.  At the very next timestep, the instantaneous total solid surface density is then redistributed into the different size bins once again. The dust grain size distribution therefore does not evolve in this model. The removal of gas with only a small mass fraction of the dust carried along results in a decrease in the gas/dust mass ratio of the remaining disk material, primarily in the inner and outer disk regions where the gas surface density decreases by the greatest factors (Figure~\ref{ape}). In earlier phases of evolution, and especially in the inner regions where viscous timescales are short, viscosity dominates photoevaporation and radially mixes regions of various gas/dust ratios to make this quantity fairly constant. At late epochs, when photoevaporation dominates, the gas/dust mass ratio is lower in regions of enhanced photoevaporation. The decreasing gas/dust ratio does not appreciably alter the disk lifetime  compared to the reference model.  This can be understood by considering the discussion above on the insensitivity of the lifetime to large changes in $\sigma_H$, since the main effect of reducing the gas/dust ratio in terms of the underlying physics is that $\sigma_H$ is increased. 

The achieved reduction in the gas/dust ratio is significant for  planet formation, as  we discuss later in \S 5.3. At $t\sim1.1-1.9$ Myrs when the gas/dust ratio at $r<10$AU has become significantly lower by a factor of $\sim 2-10$, there is still about 
 $2-13$ M$_J$ of gas in the disk. This gas is dispersed in $\lesssim$ 1Myr.  Approximately half of the total initial dust and gas advects inwards  and is lost to the star. While the remaining $\sim$50\% of the gas is removed by photoevaporation, only $\sim 10^{-4}$\ms\ or about 17\% of the initial  dust mass is carried away with the flow. Therefore, $\sim 33$\% of the initial dust mass (or $3.25\times 10^{-4}$\ms) is retained after dispersal.

\subsection{Dust Size Evolution and Radial Drift}

\begin{figure}
\centering
\plottwo{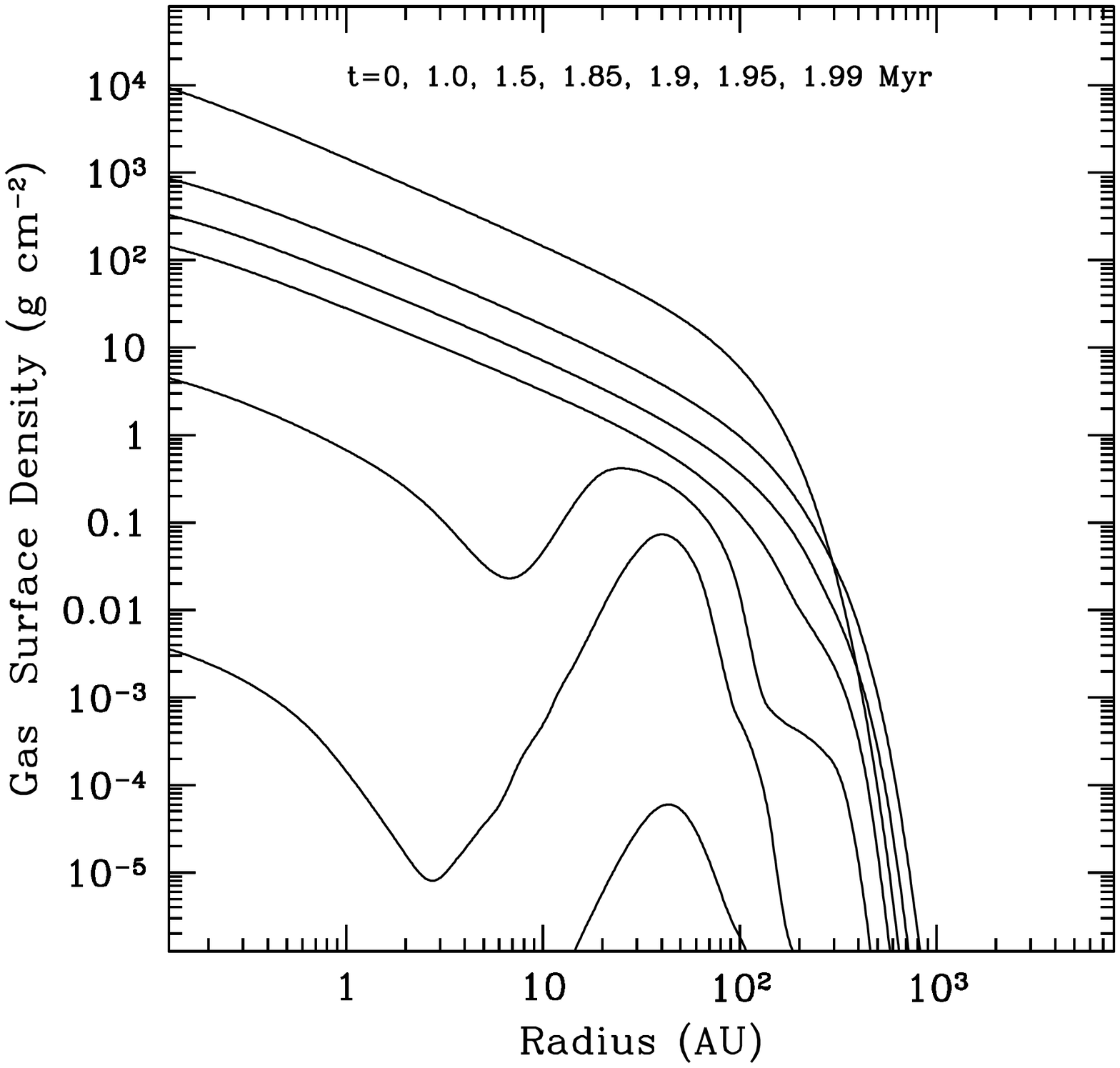}{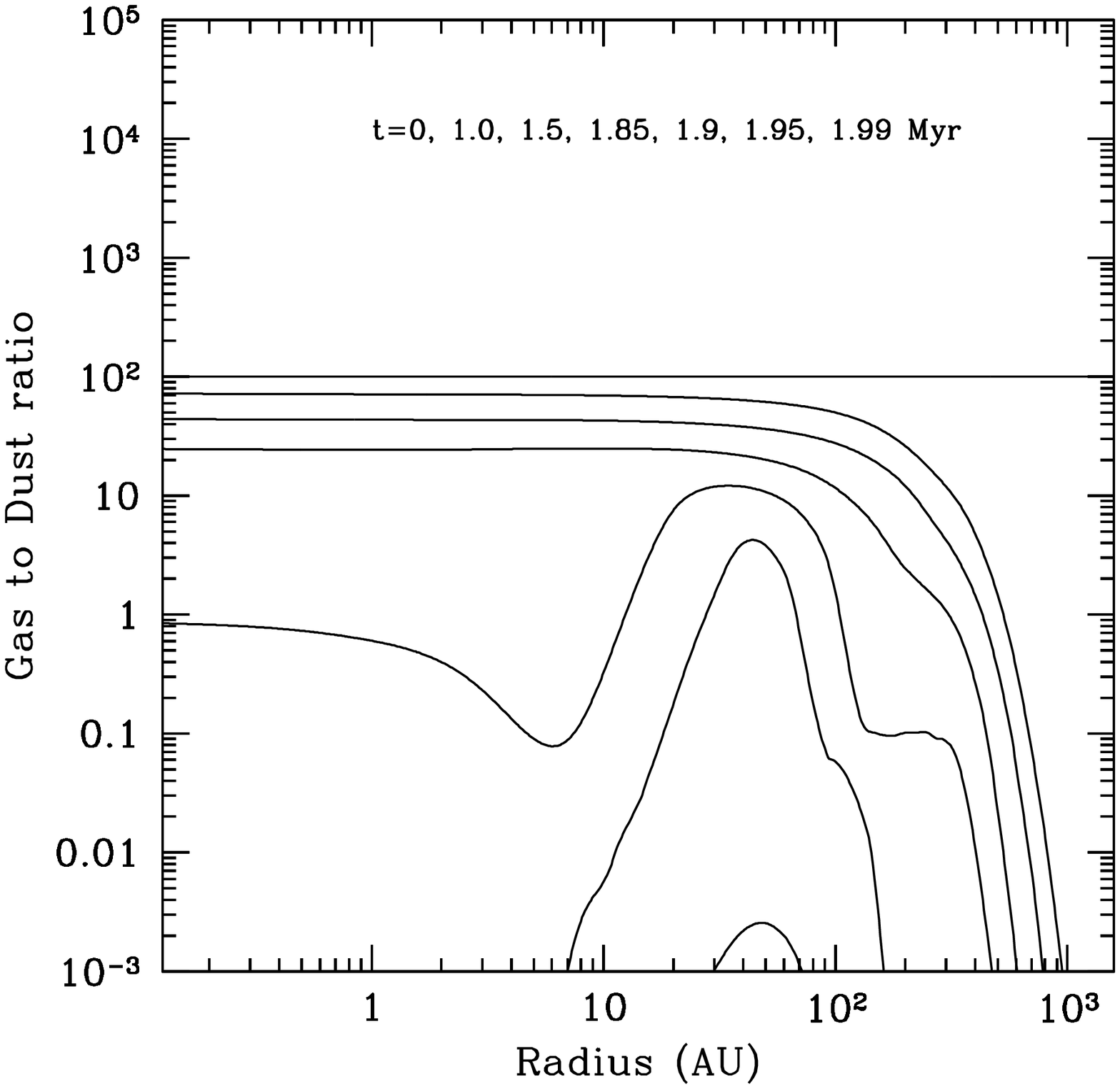}
\caption{\it Results for a model with dust evolution (BOD11) implemented; left panel shows the surface density evolution, while the right panel shows the evolution of the gas/dust mass ratio in the disk. The disk lifetime is reduced by a factor of $\sim 2$ for the chosen input parameters, but qualitatively disk evolution is similar to the reference model. The gas/dust mass ratio drops more rapidly than the case with no dust evolution.}
\label{dustevol}
\end{figure}

\paragraph{Dust evolution model}
We now consider dust evolution according to the scheme of BOD11, which includes fragmentation and growth in a turbulent disk. We assume a fragmentation threshold velocity of 10 m s $^{-1}$, above which all collisions lead to fragmentation (e.g., Beitz et al. 2011).  (We also ran a case with a threshold velocity of 1 m s$^{-1}$; disk evolution is qualitatively similar.) The level of turbulence is specified through the viscosity parameter $\alpha$. Figure~\ref{dustevol} shows the time evolution of the surface density distribution and the gas/dust mass ratio for this case. Disk lifetimes are a factor  $\gtrsim2$ lower than our reference disk model. 

Evolution of dust marginally impacts disk dispersal by enhancing FUV photoevaporation. The shallow radial temperature gradient and low surface densities in the outer disk (which increase grain-grain collisional velocities) make shattering more efficient ($a_{max}\sim a_{frag}\propto \Sigma_g/c_s^2$; Eq.~\ref{afrag}) there and limit grain growth to a few tens of microns in size. Although $a_{BT}$  is only weakly dependent on $\Sigma_g$ (Eq.~\ref{abt}), it nevertheless decreases enough to skew the grain size distribution toward small grains (for a given total dust surface density) in the outer disk as $\Sigma_g$ drops.  Thus, the  effective cross-section $\sigma_H$ increases with time to make FUV photoevaporation slightly stronger in the outer disk. As the disk continues to evolve, the  gas/dust ratio decreases to further accelerate this effect, resulting in shorter disk lifetimes (Fig.~\ref{dustevol}).   

The gas/dust mass ratio is markedly affected in the inner and outer disks due to removal of gas by photoevaporation.   Higher flow densities in the inner disk lead to greater coupling between dust and gas and relatively more dust is lost when compared to the outer disk where flow densities are low. The gas/dust mass ratio is also lower by a factor of $\sim 2-3$ from the initial value of 100 at intermediate radii ($\sim 10-100$AU) for much of the disk evolution due to both advection of low gas/dust material and photoevaporation. 
 When the gas/dust ratio falls by 2 at radii $r<10$AU in the disk at $t=1.24$ Myrs, the mass of gas left is $\sim 10$M$_J$; when the ratio falls by 10 at $t=1.95$ Myrs, the mass of gas left is $\sim 1$M$_J$.  About  $\sim 55$\% of the gas mass is photoevaporated and the rest is accreted.  After gas disk dispersal at $t=2.2$ Myrs, a dust mass of $2.9\times 10^{-4}$ \ms\ remains. There is less dust remaining from the case without dust evolution discussed earlier; this is because collisions generate smaller dust as $\Sigma_g$ declines and more dust gets entrained in the photoevaporative flow.

\paragraph{Dust evolution with radial drift}
We finally include radial drift of dust particles, which further re-distributes the dust populations as the disk evolves.  Radial drift is mainly caused by the headwind experienced by dust particles.  Gas orbits at sub-Keplerian speeds because of  radial pressure gradients. Smaller particles are coupled with the gas and have similar radial and orbital velocities, while larger particles move on almost Keplerian orbits (see Eq.~3).  Intermediate-sized particles 
experience stronger headwinds than smaller particles but have less orbital momentum than the larger particles, and therefore spiral in faster to the star, drifting with respect to the gas and small dust. Drift is highest for particles with Stokes parameter of order unity (e.g., Johansen et al. 2014). $St=1$ yields a size $a_{drift}=2\Sigma_g/(\pi \rho_s)$. Since dust evolves in the background of the changing gas surface density distribution, the size of the particle subject to maximum drift changes with $\Sigma_g$ in radius and time. 
 
\begin{figure}
\centering
\plottwo{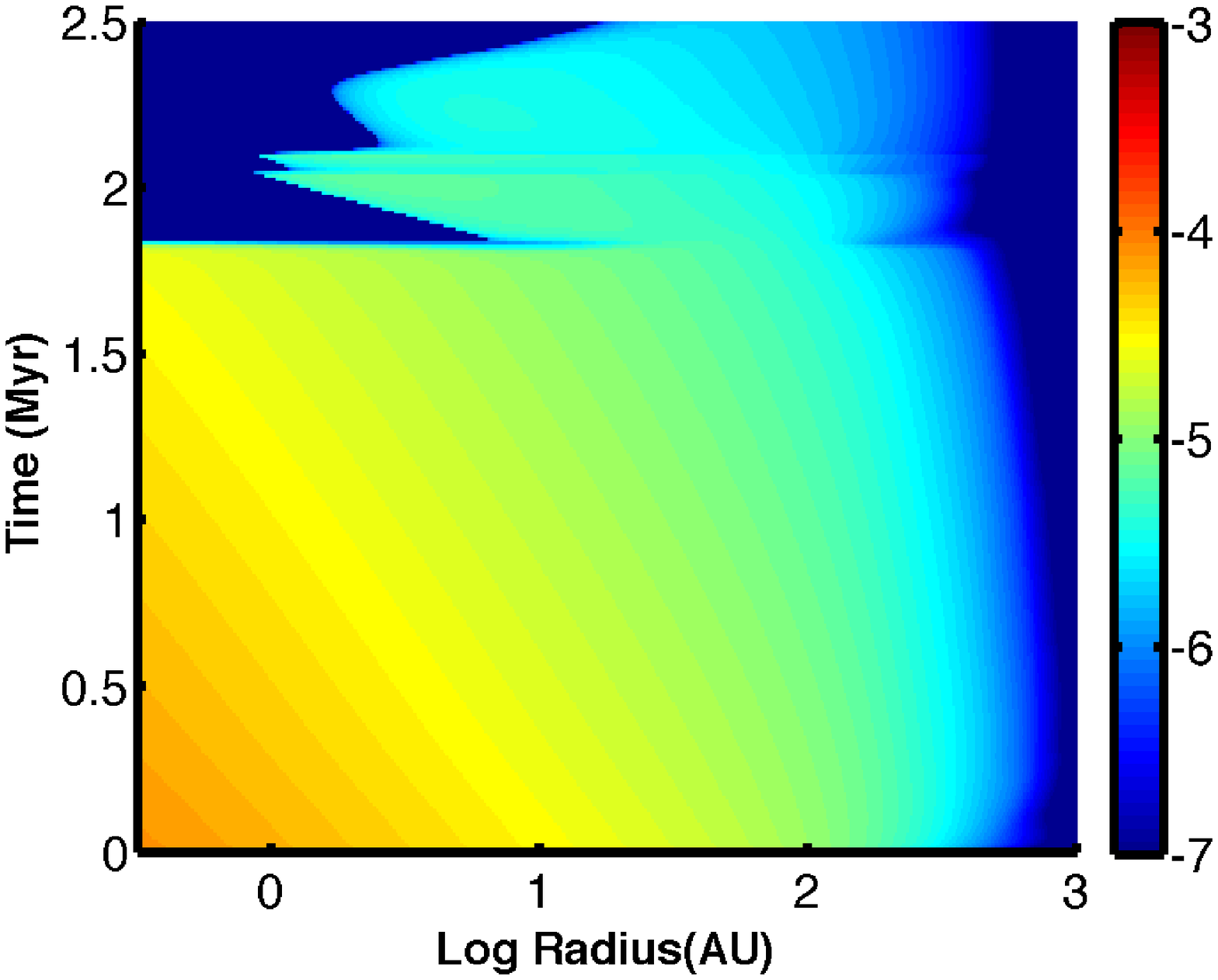}{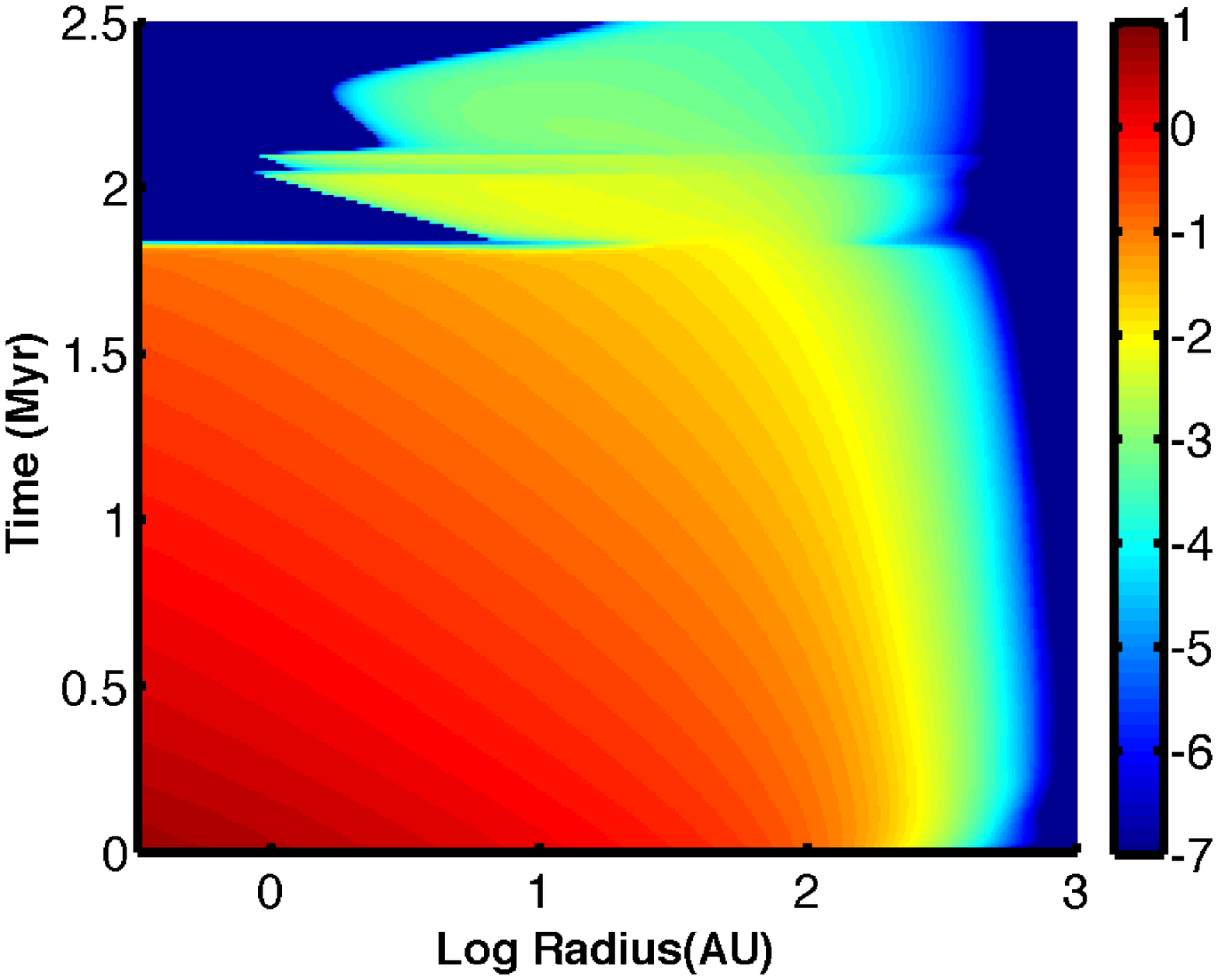}
\caption{\it  The evolution of the two critical grain sizes (in log cm units) is shown in the two panels. On the left is a color map of the size $a_{BT}$, which dominates the dust cross-sectional area, with radius and time. Initially, $a_{BT}$ is $\sim 0.3- 1\mu$m at $r<10$AU,  but over most of the disk extent it remains $\sim 0.1-0.3\mu$m until the disk disperses. The dark blue areas in the figure reflect low disk surface densities, where turbulent shattering results in small grains. Note the development of a gap prominent at $t\sim1.8$Myr at $r\sim10$AU.  The rim of the outer disk is at $\sim 10$AU when it forms, but diffuses inward to $\sim$1AU when the production of accretion-generated FUV is halted and photoevaporation rates decrease slightly. The right panel shows the maximum grain size $a_{frag}$ which is typically in the submillimeter range, and reaches $\sim 1$cm only in the innermost regions of the disk at early epochs when the surface density is high. (Note that although $a_{frag}$ evaluates to be $>1$cm in the inner disk initially, we always keep the maximum size $a_{max}$ fixed at 1cm). After gap formation, right before the disk disperses, the maximum grain size attained over the entire disk is only $\sim100\mu$m.
 }
\label{dustas}
\end{figure}

 As discussed earlier, with dust evolution $a_{BT}$ and $a_{frag}$ change with radius and time, affecting $\sigma_H$.  Figure~\ref{dustas} shows the radial and temporal evolution of  $a_{BT}$ and $a_{frag}$ for the disk model with drift; the cross-sectional area ($a_{BT}$) is dominated by grains $\sim 0.1-0.3\mu$m in size throughout the disk lifetime while the dust mass ($a_{frag}$) is in grains of size $\sim 0.01-1$cm. At a given radius, both these quantities decrease with time as the surface density decreases.  (Note the distinction between the maximum size $a_{frag}$  allowed by the collision model and the maximum size allowed in our simulations, $a_{max}$, which has an upper limit of 1cm.  Grains of sizes up to $\sim 10$ cm may be possible at small disk radii, although we redistribute this mass into 1cm particles in model calculations.)  The maximum Stokes number that can be reached in the collisional dust model is $St_{max} = u_f^2/(\alpha c_s^2)$. For our choices\footnote{Material strengths of dust particles are generally believed to be insufficient to withstand collisions to much higher velocities without fragmentation. If $\alpha$ in the disk is smaller, loss due to drift becomes more important.} of $u_f=10$m s$^{-1}$ and $\alpha=0.01$, we have the maximum Stokes number at any radius as $St_{max}(r)\sim 3/T(r)$ where $T(r)\sim 250  (r/1AU)^{-0.5}$K is the midplane temperature. The largest values of $St_{max}$ are in the outer disk, where the temperature is the lowest. For example, at 100AU, $T=25$K and $St_{max}\sim 0.12$.  At smaller radii, $T$ is greater and $St_{max}$ correspondingly smaller.  The largest particles at any given radius and epoch are therefore always such that their Stokes numbers are less than unity and these are also the particles that experience the maximum radial drift. 

\begin{figure}
\centering
\plottwo{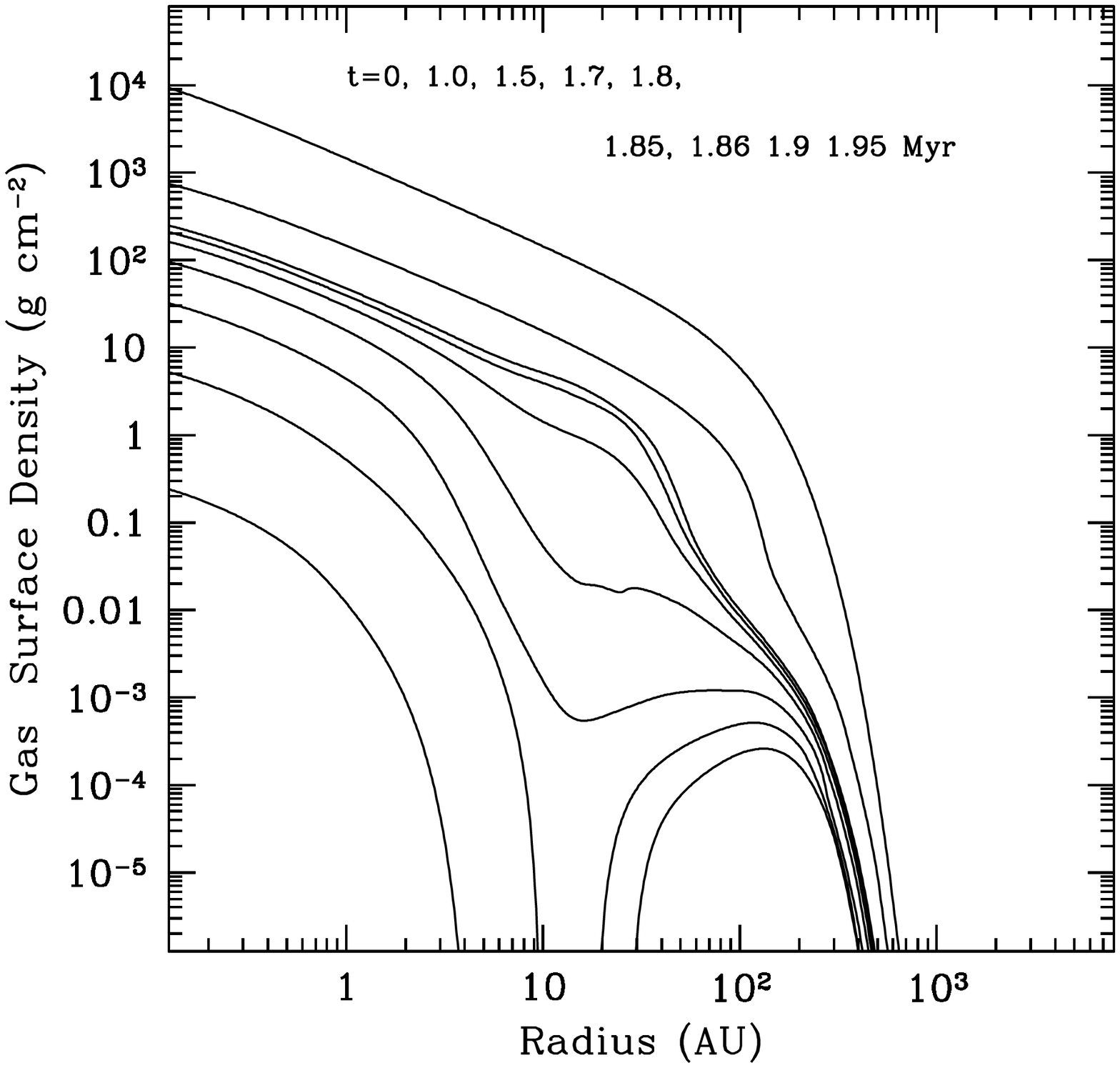}{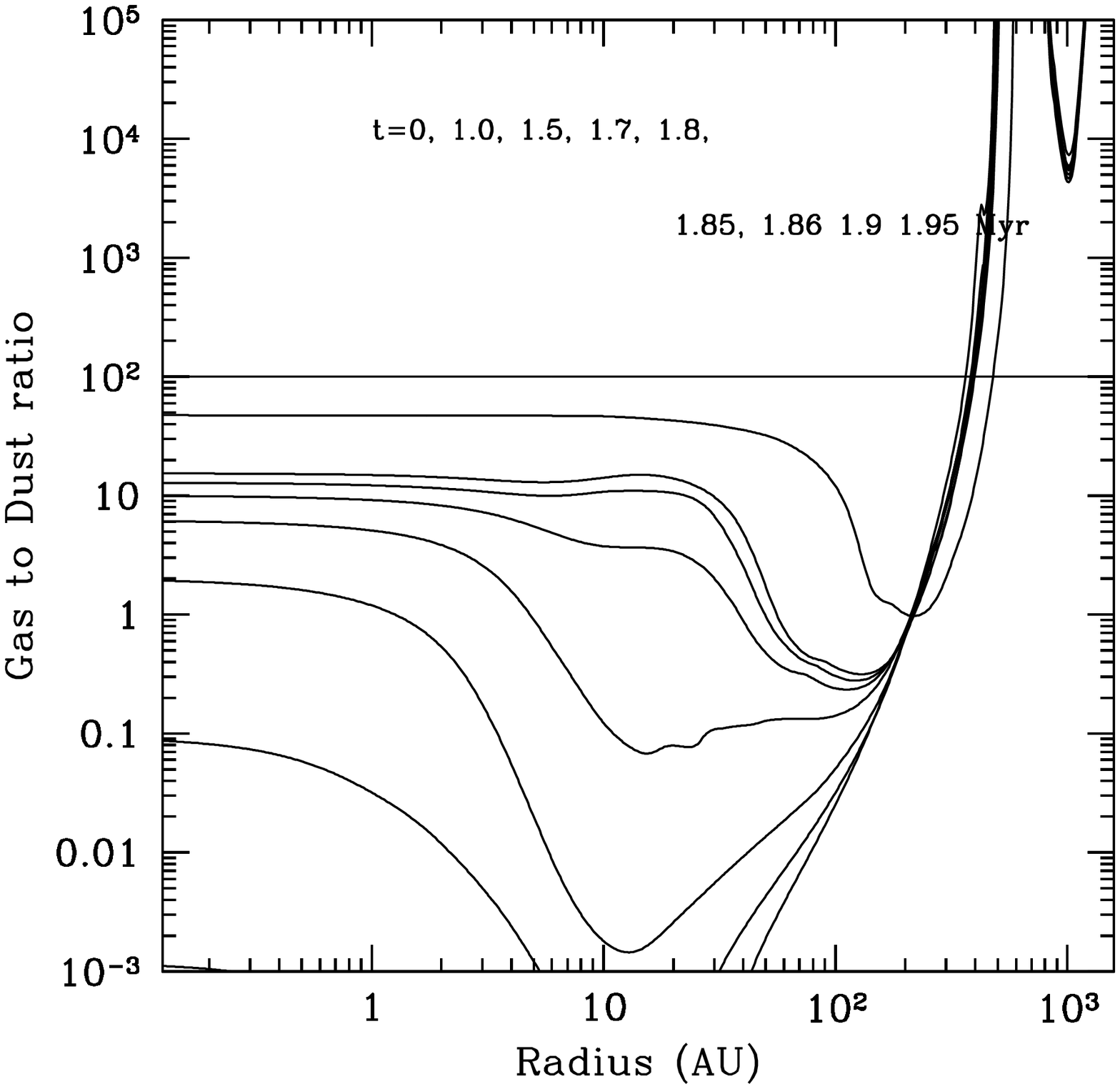}
\caption{ \it The full disk evolution model with radial drift included is shown in this figure. The introduction of radial drift of dust grains is seen to have little impact on the lifetimes of the disk when compared to the no drift model in Fig.~\ref{dustevol}. The evolution of the surface density is quite similar, apart from minor differences in the distribution of gas mass at later stages. The gas/dust ratio is however, noticeably different. Radial drift brings in dust from the outer disk, depletes the outer regions of dust and the gas/dust mass ratio rises significantly at large radii, contrary to the model in Fig.~\ref{dustevol}. }
\label{drift}
\end{figure}

Including drift for an $\alpha=0.01$ viscous model does not affect photoevaporative disk lifetimes (Figure~\ref{drift}, compare Fig~\ref{dustevol}). As discussed above, drift is only relevant for large grains.  Photoevaporation rates are sensitive to the small dust population, but this remains essentially unchanged from the case without drift. 

\begin{figure}
\centering
\plotone{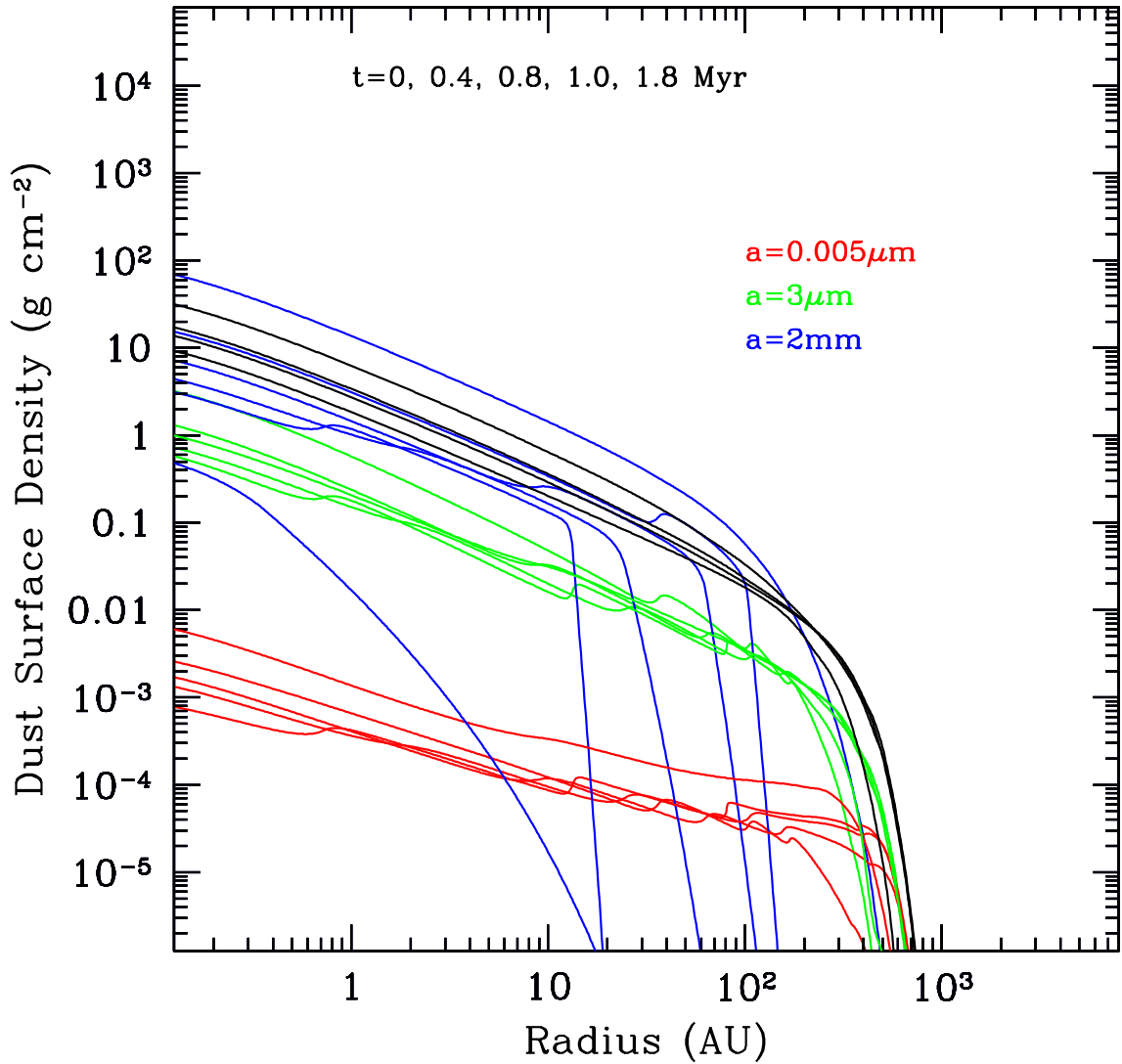}
\caption{ \it The evolution of the dust surface density for three size bins is shown for a model with a fragmentation threshold velocity of 10 m s$^{-1}$. In red, is the surface density of the smallest grains which is low at early times but increases as the critical grain radius (at which turbulence dominates) gets smaller with decreasing gas surface density. The intermediate-sized $\sim3\mu$m grains (green)  are coupled to the flow and photoevaporate with it and, moreover, do not experience significant drift. Larger, 2-mm grains (blue) need high densities to form and high column densities to survive turbulent shattering,   and are more abundant in the inner disk. These grains drift radially inward and get depleted in the outer disk.  }
\label{sigmad}
\end{figure}

Radial drift, however, does change the distribution of dust particles. The most pronounced effect of drift is in the time evolution of the gas/dust mass ratio (Figure~\ref{drift}), which increases at outermost disk radii when compared to the no drift model of Fig.~\ref{dustevol}.  We can estimate the timescale for the largest particles (with $St=St_{max}$) to drift from the outer regions. All other particles drift more slowly. The second term in Eq.~\ref{vd} can be re-written as 
\begin{equation}
v_d (r)  \sim St_{max} \frac{c_s^2}{v_{K}}\frac{d\ln P}{d\ln r}
\end{equation}
and $v_d(r)$ evaluates to $\sim 10\ St_{max}$ m s$^{-1}$ for our disk parameters at all radii. The drift time is $t_{drift}\sim r/v_d \sim 0.5$ Myrs for $St=St_{max}$ particles at 100AU where most of the mass is located. However, these particles are reasonably well coupled to the midplane gas as well (first term of Eq.~\ref{vd}) which moves outward at this radius. This slightly increases $t_{drift}$  avoiding a catastrophic loss of dust mass due to drift. Further, since $c_s$ increases at smaller radii, $St_{max}$ decreases  and the drift speed is accordingly reduced as dust drifts inward. Nevertheless, large grains drift in on relatively short timescales compared to the lifetime of the disk. It is the outermost regions that are most affected as large grains (and hence most of the dust mass) drift radially inwards,  resulting in an increase in the gas/dust mass ratio locally (Fig.~\ref{drift}). Figure~\ref{sigmad} shows the radial surface density profiles of dust  at a few epochs for three dust sizes (to maintain clarity in the plot). Because the dust sizes are re-generated at every timestep based on local disk conditions, the smaller grains which dominate the cross-sectional area grow in relative abundance as $\Sigma_g$ decreases with time ($a_{BT}$ correlates with $\Sigma_g(r,t)$, Eq.~\ref{abt}, see Fig.~\ref{dustas}). Intermediate micron-sized grains are neither small enough to be near this critical size boundary, nor are they large enough to be subject to significant drift. Large grains (mm-sized or bigger) drift radially inwards and are also preferentially formed in the inner disk where surface densities are higher ($a_{max}\propto \Sigma_g(r,t)$) and become progressively confined to the inner regions as the disk evolves (also Fig.~\ref{dustas}). Note that drift of $St=St_{max}$ particles slows down as they move inward and  $St_{max}$ decreases due to the higher temperatures. This leads to an additional decrease in the gas/dust mass ratio (compared with the case with just photoevaporation) in the inner regions as dust piles up (also see Birnstiel et al. 2012 for a similar pileup at $r\lesssim 1-2$AU due to drift but without photoevaporation). In this radial drift model, the amount of gas left in the disk after the gas/dust ratio declines by 2 at $t= 1.1$ Myrs 
is $\sim 13$M$_J$.ÊÊÊÊÊ The amount of gas left in the disk after the ratio drops by 10 at $t=1.86$ Myrs is $\sim 0.9$ M$_J$. The disk disperses at $t=2.86$ Myrs and leaves behind $2.8\times 10^{-4}$\ms\ of dust.  Most of the remaining disk mass accretes and drifts on to the star, with only about $\sim 3\times 10^{-5}$\ms\ being photoevaporated along with the gas. 

The drift of the large grains and hence most of the dust mass in toward smaller radii depletes the outer region of dust, while the declining surface density in the outer disk shifts the collisional equilibrium  dust distribution towards smaller sizes as described above. Therefore, the outer regions of the disk are expected  to be devoid of larger dust grains as the disk evolves, while gas and smaller dust extend out to larger radii. This behavior is qualitatively observed in evolved disks, for e.g., TW Hya (Andrews et al. 2012, Debes et al. 2013, Weinberger et al. 2002). This radial distribution of dust in the outer disk is mainly a consequence of the dust collisional model with viscous evolution (even in the absence of photoevaporation) and was proposed recently   by Birnstiel \& Andrews (2014) as an explanation for the discrepant gas and dust radii of observed disks at sub-millimeter wavelengths. The addition of photoevaporation slightly reduces this size discrepancy,  because mass loss in the outer disk curtails the viscous expansion of  gas here when compared to a non-photoevaporating disk. 

\subsection{Effects of PAH Abundance, X-rays and Viscosity on Disk Evolution with Dust Coagulation/Fragmentation} Having progressively developed the full disk model including all important aspects of dust evolution, we now consider the effects of varying some of the key input parameters to the models. To recapitulate, we have established that within the framework of our photoevaporation models, the gas/dust mass ratio in the disk evolves with time and is lowered in the planet-forming regions of the inner disk.  With our fixed $\alpha$-viscosity prescription, dust evolution only appears to affect disk lifetimes when there is a large change in $\sigma_H$, the effective cross-sectional area per H atom in the disk.  The PAH or small dust population, X-ray luminosity of the central star and the value of the viscous parameter $\alpha$  are other known parameters that might affect disk lifetimes (see GDH09) and gas/dust mass ratios. We examine each of these in detail. Note that in all of the following models, we use the full disk model with dust evolution and radial drift; all processes discussed in \S 4.3 and \S 4.4 are included.

\paragraph{PAH abundance} The assumed PAH abundance affects FUV heating and hence the photoevaporation rate. Although it is reasonable to expect that the number density distribution of grains extends downward from the smallest grains to PAH sizes (e.g., Tielens 2008), the distribution and abundance of PAHs in disks is very poorly quantified (e.g. Oliviera et al. 2010).  Dust grain growth in an MRN distribution to cm-sizes results in a depletion of the smallest grain abundance by factors of $\sim100$.  A similar depletion of PAHs may therefore be expected, and PAH abundances are inferred to be $\gtrsim10$ lower in  the few disks where they have been detected (e.g., Geers et al. 2007, Oliveira et al. 2008). However, PAH spectral signatures are often absent in solar-type stars. The lack of PAH features and the depleted abundances are also interpreted as a result of PAH destruction, possibly due to strong UV and X-rays (e.g., Siebenmorgen et al. 2010, Vicente et al. 2013). 

So far, we have assumed that PAH abundance scales with the smallest dust size. This size $a_{min}$ is typically much smaller than the turbulence threshold size $a_{BT}$ (also see Fig.~\ref{dustas}). In the framework of BOD11, the abundance of dust grains with $a_{min}<a<a_{BT}$, and hence PAHs is much reduced compared to the MRN power law with $a_{max}=1$cm. The total cross-sectional area of dust is dominated by grains of size $a_{min}$ for an MRN distribution, whereas in the BOD11 framework it is dominated by grains of size $a_{BT}$.  These dust grains are small enough to remain coupled to the photoevaporative flow and be removed with the gas, but  small grains and PAHs are continuously regenerated by collisions in the BOD11 scheme. Since the declining surface density of the gas with time favors the production of small grains, the PAH abundance can increase as the disk evolves (see $a=0.005\mu$m population in  Fig.~\ref{sigmad}), increasing FUV photoevaporation. 

\begin{figure}
\centering
\plotone{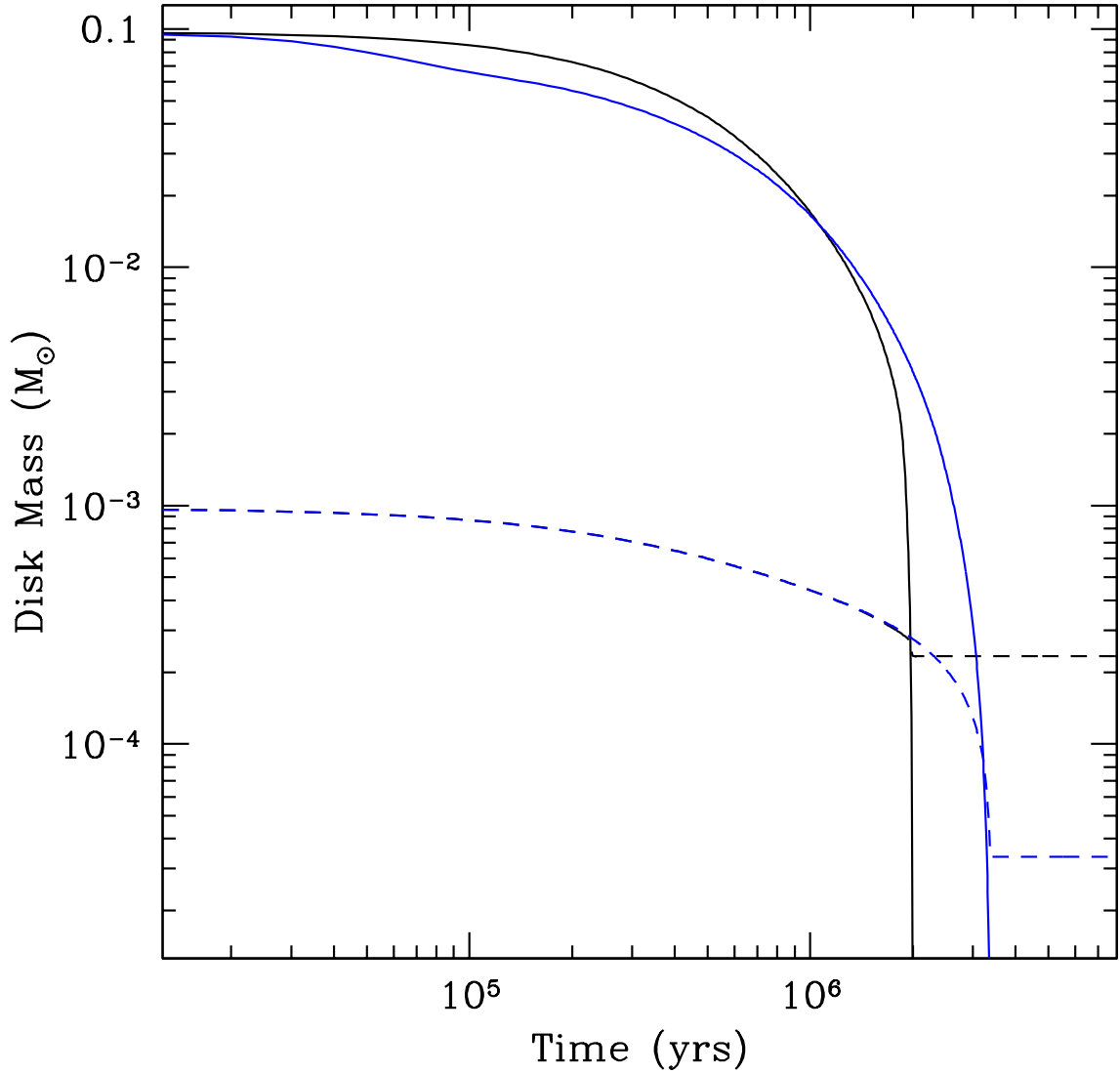}
\caption{\it Differences in disk mass evolution are shown for two models of PAH abundance. In black, the gas (solid lines) and dust mass (dashed lines)  are shown for the full dust evolution model of Fig.~\ref{drift}. A model where the PAH abundance is not scaled with the small dust population, but kept constant at an abundance $8.4\times10^{-8}$ per H atom, is shown in blue.    When the PAH fraction is scaled with the small dust, their abundance {\em relative} to gas increases as the gas/dust mass ratio decreases and this causes rapid photoevaporation. In contrast, a fixed PAH abundance model yields results more similar to the reference model, with longer lifetimes of $\sim3$ Myrs. Also note that the scaled PAH model results in a larger fraction of the dust mass being retained after gas disk dispersal. }
\label{pahs}
\end{figure}

We now consider the other extreme assumption---that PAHs are not regenerated along with the small dust grains and that their abundance remains fixed relative to the gas. With this assumption, as the disk evolves and the gas/dust mass ratio decreases with time the abundance of PAHs measured as a fraction of the gas stays constant but {\em decreases} relative to the dust.  FUV heating and photoevaporation rates thus decrease with time when compared to the scaled PAH abundance models, but at later epochs X-ray heating compensates for lowered FUV heating. Overall, disk lifetime increases as seen in Figure~\ref{pahs}, and is comparable to our ``average" scenario represented by the reference model. The gas/dust mass ratio also decreases on similar timescales, but when compared to the scaled PAH model, the fixed PAH model retains a factor of $\sim10$ lower dust mass ($\sim 3$\% of the initial value) at the end of evolution. This is mainly due to the lower rates of photoevaporation in the outer disk at later times for the fixed PAH model, which makes viscous accretion and radial drift of dust relatively more important and delays gap opening until later epochs during evolution.  

\paragraph{X-ray luminosity}
X-ray heating of disk gas is not affected by dust, and the assumed X-ray luminosity could impact our calculation of photoevaporation rates and disk lifetimes.  Although our primary interest is in FUV photoevaporation since it could be sensitive to dust evolution, we have explored the effects of changing the value of L$_X$ in order to assess the role played by X-rays. We have considered cases with lower X-ray luminosities down to $L_X=10^{28}$ erg s$^{-1}$, and find that the disk lifetime (2.2 Myrs) is not affected significantly, and the disk evolution (gas/dust ratio, surface density evolution) is essentially similar for X-ray luminosities lower than our fiducial value. 

We do not consider models with higher X-ray (or EUV) luminosities than our standard model as these cases are not relevant to dust evolution which is the focus of this paper.  Depending on the adopted spectrum (softer X-ray spectra lead to more vigorous photoevaporation, see GDH09) and the strength of the X-ray flux, mass loss due to X-ray heating may in some cases be strong  (especially see models by Ercolano et al. 2009, Owen et al. 2010, 2012) and exceed FUV photoevaporation rates. Our models typically yield lower levels of X-ray photoevaporation compared to these other models; this may be due to different assumptions of gas heating and cooling processes and treatment of X-ray attenuation. We note that although X-ray dominated photoevaporation is not significantly affected  by dust evolution, the resulting change in the gas surface density due to X-ray photoevaporation may impact dust evolution (e.g., shattering and radial drift) and gas/dust ratios (e.g., photoevaporation removing gas preferentially)  just as described in the preceding sections.  

\begin{figure}
\centering
\plottwo{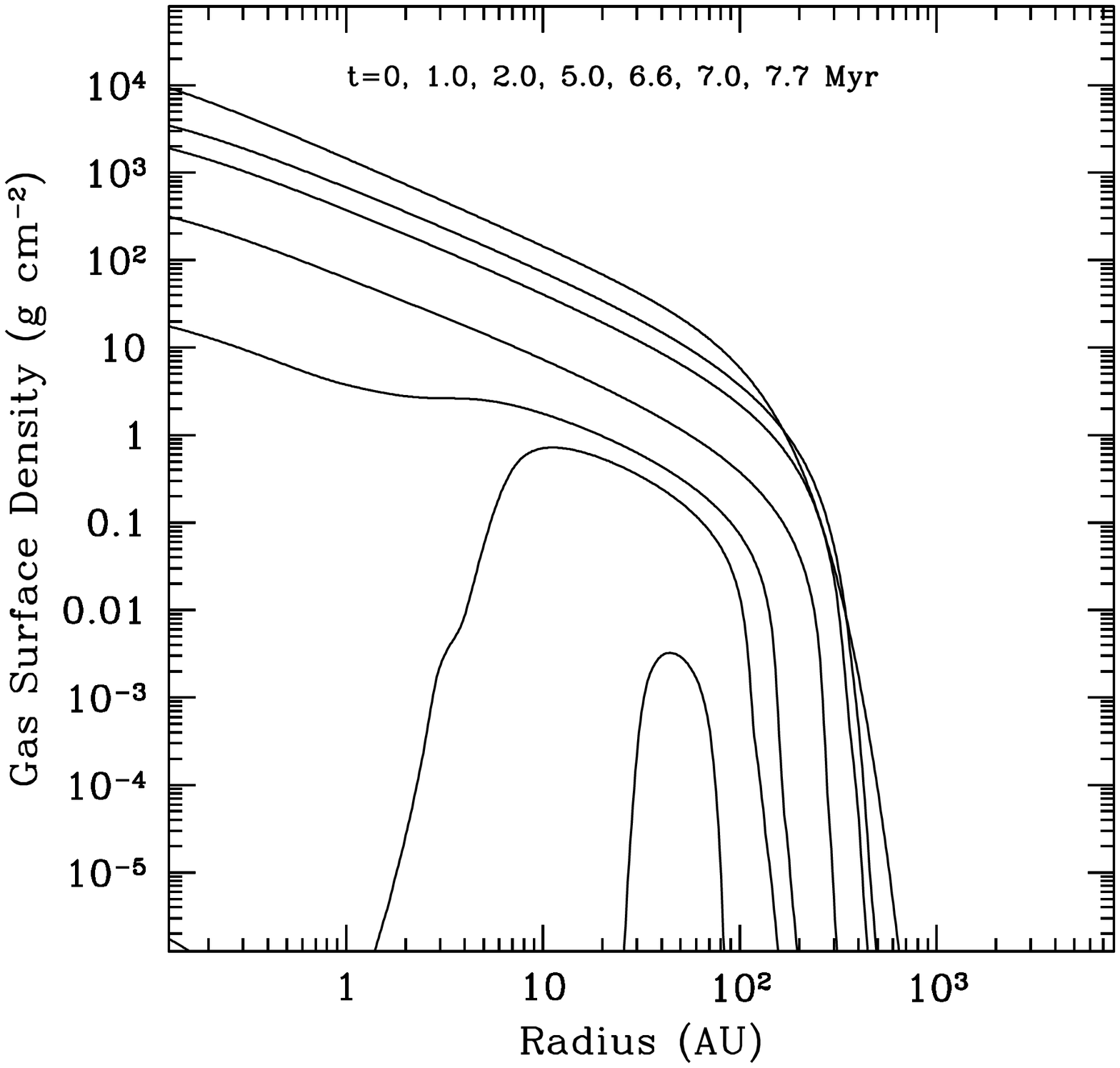}{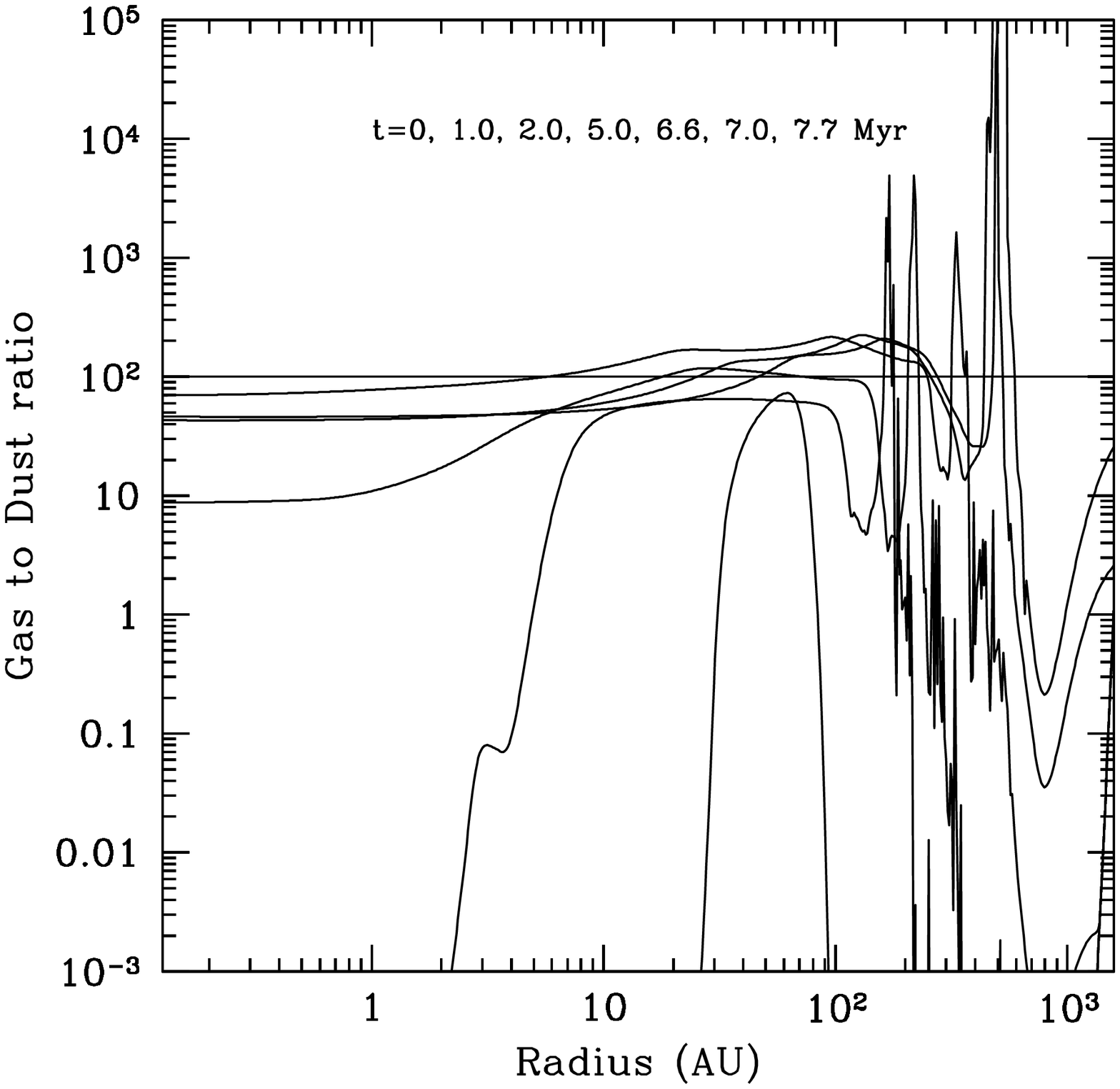}
\caption{\it The viscosity parameter $\alpha$ has a very pronounced effect on disk lifetimes. Disk lifetimes  scale nearly linearly with $\alpha$, with all other parameters fixed.  This model disk with $\alpha=0.003$ disperses in $\sim 7.8$ Myrs, compared to the $\sim 2$Myr timescale for the $\alpha=0.01$ model (Fig.~\ref{drift}). The gas/dust ratio in the disk drops to $\sim 10$ at 6.5 Myrs, when the remaining disk gas mass is $\sim 10^{-4}$ \ms.}
\label{alpha}
\end{figure}

\paragraph{Viscous parameter $\alpha$}
We find that disk evolutionary timescales are principally set by disk viscosity, which we have parametrized in a simple $\alpha$-disk prescription in our models. The adopted value of $\alpha$  influences both the nature of disk evolution and the calculated disk lifetimes. From the simple order-of-magnitude estimate in \S 4.1, the dispersal time is expected to vary as $\sim \alpha^{-1/3}$ when the photoevaporation rate itself is assumed to be independent of $\alpha$ and constant in time. However, if $\dot{M}_{pe}$ is determined by the accretion luminosity which increases with accretion rate and hence $\alpha$, the dispersal time has a stronger dependence on the level of viscosity in the disk. For example, in GDH09, we found that the dispersal time was proportional to $\alpha^{-0.6}$. In this work including dust evolution, we find an almost linear relationship---dispersal time with dust evolution is found to be proportional to $\sim \alpha^{-1.2}$ (Fig~\ref{alpha}). 

Dust collisional equilibrium and dynamics are both affected by the viscosity in the disk. The two critical grain sizes depend on $\alpha$; $a_{BT}\propto \alpha^{-0.3}$ and $a_{frag}\propto \alpha^{-1}$ as discussed in \S 3.1. Dust grains in disks with lower viscosity (less turbulence) experience less shattering and grain sizes are therefore larger. Larger grains for the same total dust surface density imply smaller $\sigma_H$ and fewer PAHs, lowering FUV photoevaporation rates. Although large changes in $\sigma_H$ are needed before the disk dispersal times change significantly (e.g., Fig.~\ref{sigmah}), low viscosity also affects dust dynamics and indirectly influences the mass loss rates. In less turbulent disks, drift is more efficient (higher $St_{max}$, Eq.~12) and more of the dust mass is lost to the star. This is reflected in the gas/dust ratio evolution in Fig.~\ref{alpha}; when compared to Fig.~\ref{drift} the gas/dust ratio is higher throughout the disk as dust is lost due to radial drift for the less viscous disk. This decrease in the amount of dust also affects $\sigma_H$ along with the increase in grain sizes---the photoevaporation rates are hence lower. Dispersal times are therefore longer for disks with lower viscosity. A disk with $\alpha=3\times10^{-3}$ dissipates in $7.8$ Myrs (Fig.~\ref{alpha}).  The gas/dust ratio in the inner regions of the disk ($\sim 1-10$AU) drops down by a factor of 2 fairly early in evolution ($t=1$ Myrs) aided by drift of dust from the outer disk. This period of gas/dust ratio being $\sim 50$ in the inner disk lasts for about $\sim 1.2$ Myrs, and later on as dust continues to drift toward the star, the gas/dust ratio rises back up again (to values greater than 100 at $r<40$ AU). Photoevaporation again causes the gas/dust ratio to drop by a factor of 2 at $t=6.4$ Myrs when the gas mass left is $\sim 1.5$M$_J$ (dust mass $\sim 1.4\times10^{-5}$\ms) and by a factor $\sim 10$ only at late times ($\sim 6.8$ Myrs), when there is approximately $\sim 0.8$M$_J$ of gas left in the disk. 
We note that an extended region with high gas/dust ratio in the outer disk is absent in this case, and hence that the radial extent of the gas and dust disks depend on $\alpha$ (as also pointed out by Birnstiel \& Andrews 2014).  Disk evolution is thus sensitive to the level of viscosity in the disk and it is hence important to consider a more sophisticated treatment of viscosity than the constant $\alpha$-parameter approach adopted here.  In future work, we will consider other  prescriptions for disk viscosity (e.g., Landry et al. 2013, Bai 2013, Okuzumi \& Ormel 2013). 

\section{Implications of Dust Grain Evolution in Disks}
\subsection{Gas/Dust Mass Ratio, Dust Retention and Disk Lifetimes}
Using a numerical disk model, we investigated disk lifetimes due to a combination of viscous evolution, photoevaporation due to EUV, FUV and X-rays, and a collisional growth/fragmentation equilibrium framework for dust evolution. Depending on model assumptions, we find that including dust evolution may lower disk lifetimes by a factor $\lesssim 2$ due to enhanced FUV photoevaporation. The shorter lifetime is mainly due to the fact that the dust collisional model in an evolving gas disk results in an increase in $\sigma_H$ with time as the gas surface density decreases.  A more robust result---independent of many disk parameters---is that the dust mass relative to the gas is enhanced in the disk as it evolves. This is primarily because very little dust is coupled to the escaping gas in photoevaporative flows. The enhancement of dust mass compared to gas mass depends on the relative amounts of material lost to photoevaporation versus accretion (which does not in general change the gas/dust ratio). In fact, when all the gas is dispersed at the end of the disk evolution, our models typically find that a substantial mass of dust still remains. Presumably, if we had included planetesimal formation, this dust would provide the necessary mass reservoir. In addition, as giant planets grow the gas and dust mixture they accrete will be enhanced in heavy element  abundances (e.g. Guillot \& Hueso 2006).  To the extent that volatile elements such as C, N and O are incorporated in grains, these will also be enhanced in abundance. 

Combined viscous-evolution/photoevaporation models have been studied by various authors (Takeuchi, Clarke \& Lin 2005, Alexander \& Armitage 2007, Hughes \& Armitage 2012), but these authors all consider photoevaporation only by EUV photons. They find, in contrast to this work, that although there can be some local pile-ups with higher dust concentrations, no significant enhancements  or reductions in the gas/dust ratio result in their models.  The main difference with our work can be attributed to the nature of EUV photoevaporation versus FUV photoevaporation as considered here. EUV photoevaporation rates are low  and dispersal is rapid only after a gap is opened in the disk at late stages. The lower rates of dispersal imply that the disk loses most of its initial mass by accretion onto the central star and must also similarly lose the dust. Indeed, Hughes \& Armitage (2012) conclude that most of the dust is lost and that viscosity smears out any local enhancements in dust at late times which were reported by previous authors. They find that the mass of solids retained in the disk is
what remains in the disk after gap opening and is  a very small fraction of initial dust mass. 

Our results differ from these previous studies primarily because FUV (and X-ray) photoevaporation rates are typically higher than EUV photoevaporation rates. FUV photoevaporation notably depletes gas in the outer region of the disk. The declining gas surface density of the disk due to FUV photoevaporation and viscous spreading decreases gas/dust coupling and there is reduced radial transport of dust by the accreting gas. Since the outer disk forms the mass reservoir, this results in a significant retention of dust mass as the disk continues to disperse. Moreover, the high mass loss rates imply that photoevaporation begins to dominate viscosity at earlier stages of disk evolution.  Gaps are thus formed when the disk is relatively massive---in contrast to pure EUV photoevaporation---and the outer disk still has a significant fraction of its initial mass when rim erosion takes over. At this stage, nearly all the dust in this outer disk is retained since it does not get carried away with the photoevaporative flow. Since almost half the initial gas mass is lost by photoevaporation in our models, a commensurate amount of dust ($\sim 30$\%) is left behind---this may then eventually form planets.
 
As mentioned above, including a dust evolution model does not significantly change calculated disk lifetimes when compared to the assumption of a dust size distribution constant with radius and time. The main reason for this result is that  the disk dispersal time and FUV photoevaporation rate are very weakly dependent on the changing collisional cross-section $\sigma_H$ as dust evolves. As FUV photoevaporation removes gas and enhances the dust/gas ratio, larger amounts of small dust per H are produced which increase FUV heating with time. However, these dust grains also decrease the penetration depth of FUV photons so that they are absorbed in lower density regions closer to the surface. These two opposing effects result in a mass loss rate that only increases slightly with $\sigma_H$.  While dust evolution and the gas/dust ratio appear to in turn be considerably affected by the evolving surface density distribution due to viscosity (which affects the grain size distribution, BOD11) and the selective photoevaporation of gas (and small dust), disk lifetimes only change by a factor of $\lesssim 2$ when a detailed dust evolution model is considered. 

\subsection{Dust Observations}
Dust continuum emission from disks, ranging from the near-infrared to radio wavelengths, still remains the main probe of disk evolution, even though gas is believed to be the dominant mass constituent through most of the disk's lifetime (see recent reviews, e.g., Williams \& Cieza 2012,  Testi et al. 2014). Disk mass evolution due to photoevaporation and viscous evolution as described here may have an observable outcome on dust emission. The consequences of dust evolution alone  for disk emission have been explored in great detail by several studies that use the state-of-the-art dust models of BOD11 considered in this work (Birnstiel et al. 2010, 2012, Birnstiel \& Andrews 2014, Pinilla et al. 2012). In earlier sections, we have shown that photoevaporation changes the disk surface density profile and its evolution, thereby affecting the evolution of dust through several parameters that all depend on $\Sigma_g$: the maximum size $a_{frag}$, the cross-sectional area of grains via $a_{BT}$, the Stokes coupling parameter $St$ for a dust particle of given size and the resulting diffusivity. Therefore, photoevaporation may affect many  of the results previously reported:  early rapid growth of dust particles to large sizes,  different observed radial sizes of gas and dust as the disk evolves, and even the inferred evolution of dust in the disk. We discuss these below.

The extent to which grain growth has occurred in a disk is determined by the slope of the submillimeter emission ($\beta$) which declines if grains larger than about 0.1 mm are present with substantial mass (e.g., Draine 2006). Measured values of $\beta$ with some certainty indicate  the growth of dust grains early in a disk's evolutionary history; grains of up to cm sizes have been inferred (e.g., Beckwith \& Sargent 1990, Natta et al. 2004). There are some young disks observed that already appear to have low millimeter fluxes and low spectral indices (e.g., Ricci et al. 2010, Ubach et al. 2012), perhaps due to the early formation of very large grains (cm-sizes) in cores and protostellar envelopes even before star formation (e.g., Laibe et al. 2013, Schnee et al. 2014, Miotello et al. 2014). The BOD11 dust evolution models however suggest an increase in $\beta$ with age; grains get smaller as the surface density decreases. With photoevaporation, surface density decreases even faster in time than pure viscous models and a stronger increase in $\beta$ is to be expected as disks age. As yet, there is no clear evidence for this increase in $\beta$, although a recent study by Pinilla et al. (2014) shows that transition disks---which are more evolved--- have relatively steeper spectral slopes than typical T Tauri disks. However, as summarized by Testi et al. (2014), tracing the evolution of dust emission with age is not straightforward, due to current uncertainties in both dust optical properties and determination of age/evolutionary status of individual disks.

  The difference in the radial sizes of disks as probed by gas and dust has long been a puzzle and  initially attributed to the use of radially truncated vs. tapering surface density profiles in modeling (e.g. Hughes et al. 2008); however, this difference was recently demonstrated as possibly a true discrepancy (Andrews et al. 2012) and explained as the result of radial drift of dust in an viscously expanding disk (Birnstiel \& Andrews 2014). Photoevaporation restricts the viscous expansion of the outer disk (see GDH09 for a more detailed discussion) and also rapidly depletes the surface density here.  Lower surface densities inhibit grain growth and result in a sharper truncation of the outer dust disk than the purely viscous case. This is shown in Figures~\ref{drift} and \ref{sigmad}, gas depletion in the outer regions limits the extent of the disks observed at mm wavelengths because grain growth to millimeter  sizes is restricted to smaller and smaller disk radii as the disk evolves. On the other hand, the abundance of small grains increases with time at large radii. Scattered light emission from small grains in the outermost regions of the disk may also be expected to be enhanced for more evolved disks.  Such effects have already been detected in the evolved disk around TW Hya (e.g., Andrews et al. 2012, Debes et al. 2013). Although the enhanced abundance of small grains in the outer disk relative to the larger dust is also present in models without photoevaporation and is due to drift (e.g., Birnstiel \& Andrews 2014), these effects are more pronounced when photoevaporation is included because of the sharper decline in gas surface density in the outer disk. As shown in \S 4.5, disk viscosity also affects the radial extent of the gas and dust disks. However, because of the smaller radial extent due to photoevaporation,   the difference in gas and sub-millimeter dust disk radii is expected to be smaller than for pure viscous disk evolution. 

Most of what we know about disk evolution has come from observations of transition disks (Williams \& Cieza 2011). Disks are believed to evolve along several distinct evolutionary pathways, with some disks developing holes or gaps while still massive --- some of these accrete while others do not.  ``Anemic" disks are also inferred to exist, these show a steady decline in their infrared fluxes and no indications of gaps or holes (e.g., Currie et al. 2009, Sicilia-Aguilar et al. 2010, Williams \& Cieza 2011).  Some of these scenarios may be explained in the context of disk evolution as described here. Transition disks with holes and no accretion can be clearly interpreted as disks that may have evolved past the gap opening epoch in a photoevaporating disk (e.g., Williams \& Cieza 2011 and references therein). Many transition disks, however, show signs of accretion and appear to be incompatible with holes having being carved out by photoevaporation. Planets are believed to be responsible for these disk cavities, but their formation could have been aided by photoevaporation as will be discussed shortly. In the inner disk large grains can form even at late epochs due to higher densities. If grain growth is extended to even larger objects (current models are restricted to cm-sized grains), or alternately if the decreasing gas/dust mass ratio triggers the growth of planetesimals and consequently, rocky planets and cores of giant planets, the opacity in the inner disk may be significantly reduced, perhaps explaining the occurrence of at least some accreting, transition-type disks.  Anemic disks may be the result of low chromospheric luminosities of the central sources and a combination of viscous evolution and photoevaporation (see the ``No Gap" model of Fig.~\ref{overview}). The prolonged presence of gas in these inner disks and the lack of holes isolating the outer disk may lead to most of the dust mass being eventually accreted onto the star, rather than being left behind in a disk cleared of gas by photoevaporation. 

\subsection{Planet Formation}

The most interesting implications of photoevaporation with dust evolution relate to the evolution in the gas/dust ratio and consequences for planet formation. Many direct routes to planet formation have been invoked to circumvent the numerous obstacles to planetesimal growth (e.g., bouncing barriers, fragmentation, radial drift) and subsequent planet formation (Johansen et al. 2014). Some of these include gravitational instability (Goldreich \& Ward 1974), vertical settling and sweeping/collective growth (Youdin \& Shu 2002), turbulent concentrations in eddies (Cuzzi et al. 2008, 2010) and streaming instabilities (Johansen \& Youdin 2007, Bai \& Stone 2010). Most of these processes require a reduction of $\sim2-10$ in the gas/dust ratio to operate efficiently. We find that reductions in the gas/dust ratio of the required magnitude in the planet-forming regions of the disk  are  a natural consequence of disk models that include photoevaporation and dust evolution.  
Photoevaporation removes gas and negligible amounts of dust mass, reducing the gas/dust mass ratio in the disk and affecting further dust evolution. In the BOD11 framework adopted here, dust evolution under lower gas surface density results in enhanced photoevaporation, amplifying the reductions in gas/dust ratio in the disk. Radial drift by itself without photoevaporation may provide modest reductions ($\sim$ 2) in the gas/dust mass ratio, but only  in the innermost $\lesssim 1-2$ AU regions under the right disk conditions such as a low viscosity (see Birnstiel et al. 2012). In contrast, photoevaporation reduces the gas/dust mass over a much wider region ($\sim$ tens of AU) by a considerable factor.  When the gas/dust ratio decreases to small values, any one of the above processes may operate to collect the dust into larger planetesimals. Such a mechanism for forming giant planets was proposed by Throop \& Bally (2005) in the context of external UV photoevaporation of disks.

According to core accretion theory,  giant planet formation begins with the formation of planetesimals, is followed by the formation of rocky cores of giant planets, and then concludes with the cores collecting some fraction of the gas and dust left in the disk to form gas-poor or gas-rich giant planets.  In this scenario, the gas mass available once planetesimals have formed and the time available for gas accretion before the disk disperses are key factors. ÊAs discussed above, planetesimal formation may require the reduction of the gas/dust ratio by factors $2-10.$ To achieve this reduction, photoevaporation must have removed a large fraction of the gas mass in the planet forming zone. Planet formation could thus be tied to disk dispersal. The results presented in this paper suggest that once the gas/dust ratio has dropped by 2 in the planet forming region of $1-10$ AU, there is $\sim 2-15$M$_J$ of gas left in the disk, and there is $\sim 1-2$ Myrs until disk dispersal.  Some recent studies of CO isotope emission in disks support such depleted gas/dust ratios in more evolved disks (Williams \& Best 2014), although measurements of gas mass are quite uncertain and depend on disk chemistry and modeling (e.g., Thi et al. 2010, Gorti et al. 2011, Bergin et al. 2013, Favre et al. 2013).
If planetesimals do not form until the gas/dust ratio has declined by 10, then we find that
only $\sim 1-2$M$_J$ of gas remains, and the time until complete dispersal of the gas is $\sim 0.5$ Myr.ÊÊ  Therefore, if planetesimals can form when the gas/dust ratio has dropped by $\sim 2$, there is sufficient gas to form Jupiters although the time available to form cores and accrete gas is somewhat short in our models.  If planetesimals can form  only when the gas/dust has dropped by $\sim 10$, the ability to form Jupiters or even massive cores of giant planets appears to be very constrained. Even if massive cores could
form on  relatively short timescales, there is not much gas available for accretion and very little time for accretion onto the core.Ê  Such a scenario would favor the formation of Neptunes and mini-Neptunes (e.g., Movshovitz et al. 2010, Rogers et al. 2011).Ê Our models 
thus suggest an explanation for why planet formation timescales and disk 
dispersal timescales are connected, and why gas-poor giants may be more common 
than gas-rich giants.ÊÊÊÊÊ Further exploration of parameter space is warranted to 
better constrain both the time available after significant drops in gas/dust ratios are achieved and the amount of gas available at this epoch.ÊÊ In particular, it appears that a better formulation for disk viscosity may affect these results, and future work will investigate other prescriptions for viscosity.

Although not included in this study, planetesimal formation may in turn affect the final stages of disk dispersal. On the one hand, formation of planetesimals may speed up disk dispersal. Once planetesimal formation is triggered at a radial zone, this process will locally deplete the disk of  some of the ``dust" ($\lesssim $1cm size) on short timescales (e.g., Andersen et al. 2014). The efficiency with which the solids are depleted into planetesimals and the fraction of mass left as dust is unclear and the depletion factor depends on the unknown details of these processes. If, for example, all the dust were converted to planetesimals by instabilities in the inner disk after the required decrease in the gas/dust ratio, the reduction in dust opacity (assuming  gas opacity is not substantial) would allow FUV photons to more easily penetrate to outer disk regions. This would lead to enhanced irradiation of the rim at the outer radial edge of the instability. If planets form rapidly and create gaps, this could also enhance the irradiation of the outer rim.  Rim photoevaporation rates are usually roughly an order of magnitude  higher due to direct illumination (e.g., Alexander et al. 2006) and outer  disk dispersal would be even more rapid.  On the other hand, planetesimal formation may slow down disk dispersal.  Larger planetesimals when they form may sweep up small dust very efficiently (e.g., Windmark et al. 2014). A decrease in dust content may impact further photoevaporation of the disk, and in fact the  requirement of $\sim 0.5$ Myr timescales for giant planet formation discussed above may be mitigated if disk dispersal slows after planetesimal formation. 
 Such an effect was also seen in the low viscosity disk model where drift depleted the dust mass and helped in increasing the disk lifetime. Since all these effects depend sensitively on unknowns such as the critical gas/dust mass ratio at which instabilities are triggered and the efficiency of conversion of dust mass into planetesimals, it is not possible to address these issues in the present study.

We note that in the above we have assumed that giant planets form via core accretion; gas giants may also form via gravitational fragmentation of a massive disk and there are some meteoritic studies that support this formation mechanism.  Calcium-Aluminum Inclusions (CAIs) found in meteorites are the oldest objects in the solar system. CAIs are believed to have formed over $\sim 0.7$ Myr (MacPherson et al. 2012) at high ($>1350$K) temperatures in gas of solar composition (e.g., Davis et al. 2014), most likely in the innermost regions of the disk very early during evolution (Class 0/I phases). A recent study of iron meteorites  indicates that they (and hence their parent planetesimals) formed only $\sim 0.1-0.3$ Myrs after CAIs (Kruijer et al. 2014), and therefore at least in the inner solar system, some planetesimal formation may have taken place without any reduction in the gas/dust ratio due to photoevaporation. Such short formation timescales may require gravitational fragmentation of the disk when it is relatively massive. Gas giant planets could thus have formed early, although it is not clear if they can survive migration in a massive disk. Core accretion can, however, explain many properties of our solar system and most exoplanets and is believed to be the dominant mechanism, with disk fragmentation playing in a role in some exoplanetary systems (see Helled et al. 2014 for a recent review). 

 A potential scenario that might bring the photoevaporation-induced planetesimal formation in agreement with meteoritic data would be the following: CAIs could have formed over an extended period (e.g., due to shocks) after the formation of the protoplanetary disk ($\sim 1-2$ Myrs). During that time, most CAIs radially drifted into the star. The CAIs we observe in meteorites are then only the last ones formed, and therefore do not represent $t=0$, but instead a later time by which photoevaporation halts accretion, increases the dust-to-gas ratio and triggers planetesimal formation. The triggering of rapid planetesimal formation due to a reduction in the gas/dust ratio is  consistent with chondrules originating from planetesimal impacts and cooling of the resulting sprays of melt (e.g.. Kieffer 1975, Asphaug et al. 2011, Dullemond et al. 2014, Johnson et al 2015); chondrules form over a $1-2$ Myr  period after planetesimals form. However, it is quite probable that both gravitational fragmentation and core accretion are responsible for planet formation in many disks (e.g., Helled et al. 2014), and it is only for the core-accretion theory that photoevaporation may aid planetesimal formation.

Irrespective of the actual mechanism for planetesimal formation, we can conclude that a large fraction of the initial disk solids mass is retained after the disk disperses. This forms a reservoir for the formation of terrestrial planets and planetary objects. The final solid masses in our dust evolution models (with $\alpha=0.01$) are $\sim 3  \times 10^{-4}$\ms, comparable to the solids inventory of our own solar system (Chiang \& Youdin 2012). This represents about 30\% of the initial dust mass for a 0.1\ms\ disk, the rest is either accreted onto the star or lost along with the photoevaporative wind. 

\section{Summary}
We present the first models that calculate the evolution of a protoplanetary disk including the effects of viscosity, photoevaporation due to EUV, FUV and X-rays, and dust dynamics and size evolution in an evolving gas disk. Using a collision-fragmentation equilibrium scheme for dust evolution following Birnstiel et al. (2011) and a simplified treatment of gas chemistry and thermal balance, we solve for the viscous and photoevaporative evolution of these two components in a multi-fluid approach. We also consider the effects of radial drift and gas/dust coupling in the photoevaporative wind. Our main conclusions are:
\begin{enumerate}
\item Assuming EUV, X-ray and FUV luminosities are equal, and accretion generates FUV photons, the cross-sectional area presented by dust grains ($\sigma_H$) is the main determinant of the rate at which mass is lost due to FUV photoevaporation.  Higher $\sigma_H$ results in faster disk dispersal. However, the dependence of the photoevaporative mass loss rate on $\sigma_H$ is surprisingly weak. Lower values of $\sigma_H$ lead to less heating, but greater penetration of FUV photons to higher density gas provides a compensatory effect. 

\item Including a coagulation/fragmentation model for dust evolution results in an increase in the small dust population as the surface density in the disk declines with time, leading to higher turbulent speeds of dust particles and more dust shattering. Disk lifetimes with dust evolution and photoevaporation are a factor of $\lesssim 2$ smaller than our fiducial case with no dust evolution. 

\item Although dust evolution does not significantly affect disk dispersal timescales, we find that disk dispersal is important to consider in studies of dust evolution and planet formation. The evolving surface density in the disk has a considerable influence on dust growth and dynamics. 

\item The level of viscosity in the disk strongly affects disk evolution and dispersal times. In our constant-$\alpha$ viscosity model, disk lifetime is proportional to $\alpha^{-1.2}$ with dust in less viscous disks being more subject to loss by radial drift, thereby increasing the gas/dust ratio. 

\item Large reductions ($\sim 2-10$) in the gas/dust mass ratio are achieved due to photoevaporation, making conditions favorable for the formation of planetesimals by instabilities. There may be as much as $\sim 2-15$M$_J$ of gas remaining in the disk when the gas/dust ratio drops below 2 and nearly $\sim 1-2$M$_J$ of gas left when the gas/dust ratio drops below $\sim 10$. Complete dispersal is rapid after these epochs, with the disk typically dissipating in about $\sim 0.5-1$ Myrs after the gas/dust ratio drops by 10. There may thus be enough gas mass left to form gas giants; the formation process has to be rapid not only to form massive cores from planetesimals, but also  to allow gas accretion onto cores before disk dissipation.

\item   The accretion of the remaining gas/dust reservoir onto the massive cores may result in an enhancement of heavy elements in giant planets. If some fraction of C, N and O are incorporated into the dust, even these relatively volatile elements may have enhanced abundances relative to hydrogen. 

\item A large fraction of the dust mass in the disk is typically retained after the disk has dispersed. FUV-driven photoevaporation in the outer disk, where most of the disk mass lies, is mainly responsible for this effect which is due to the impact of the rapid drop in gas surface density at large radii on dust growth and radial transport.  High photoevaporation rates result  in a significant fraction of the disk gas being preferentially removed. Gaps form while the disk is still massive and aid retention of the nearly all the dust in the outer disk. This result differs from previous cases which only included EUV photoevaporation. We conclude that remnant dust could provide the mass needed for the formation and growth of planetesimals as the disk disperses. We find that viscosity and drift deplete $\sim 55$\% of the dust mass onto the central star, $\lesssim 15$\% is photoevaporated, and $\sim 30\%$ of the total initial dust mass remains after disk dissipation.
\end{enumerate}

We end with the following caveats on important processes that are likely to affect the above conclusions. The main uncertainty in the models pertains to the poorly understood nature of disk viscosity; future work will consider more physically motivated prescriptions to include its effects. Additional mass loss may occur due to magneto-centrifugal disk winds, especially in the inner, $\lesssim 10$AU regions where they operate concurrently with disk accretion. These winds may be denser (e.g., Bai 2013) than the thermal, photoevaporative winds and may significantly deplete the inner disk of dust. Although disk winds carry mass at only a fraction of the accretion rate and hence cannot dominate accretion, they may still enhance some of the results of photoevaporation presented here.  Magnetic disk winds are, however, beyond the scope of the present work.  A last issue is the unknown evolution of small grains and PAHs in disks around young, low mass stars. They could be ablated or at least partially destroyed by X-rays and UV photons (e.g., Pety et al. 2005, Siebenmorgen et al. 2010), or even altered by chemical evolution (e.g., Cook et al. 2014). These processes are the subject of current laboratory and theoretical investigations and we soon hope to have sufficient information to include their effects in future studies. 
\acknowledgements
Uma Gorti acknowledges several helpful discussions with Jeff Cuzzi, Chris Ormel and Til Birnstiel. U.G and D.H acknowledge funding from NASA Award NNX09AO42G which made this work possible. This project made use of NASA HEC supercomputing resources. 
C.P.D acknowledges funding from DFG grant DU 414/9-1.

\begin{table}

\caption{\it Default model parameters}
\begin{tabular}{ll}
\\
\hline
Model parameters \\
\hline
Stellar mass & 1\ms \\
Initial disk mass & 0.1\ms \\
L$_X$ & $2\times10^{30}$ erg s$^{-1}$ \\
L$_{FUV}$ {\small (Chromospheric component)}  & $2\times10^{30}$ erg s$^{-1}$ \\
L$_{EUV}$ & $2\times10^{30}$ erg s$^{-1}$ ($\Phi_{EUV}=10^{41}$ s$^{-1}$) \\
Dust grain size & 0.005$\mu$m $<a< 1$cm\\
Gas/dust mass ratio  & 100 \\
Viscous parameter $\alpha$ & $10^{-2}$ \\
\hline
\end{tabular}
\end{table}

 \end{document}